\documentclass[preprint,aps]{revtex4}
\usepackage{epsfig}
\usepackage[dvips]{color}

\begin{document}

\title{ \Large  Correlation regimes in fluctuations\\ of
fatigue crack growth}
\author{  Nicola Scafetta$^{1}$, Asok Ray$^{2}$, Bruce J. West$^{1,3}$}

\address{$^{1}$Department of Physics, Duke University, Durham, NC 27708}
\address{$^{2}$ Mechanical Engineering Department, The Pennsylvania State University,
University Park, PA 16802}
\address    {$^{3}$ Mathematics Division, Army Research Office, Research Triangle Park, NC 27709 }

\date{\today}

\begin{abstract}
 This paper investigates correlation properties of
fluctuations in fatigue crack growth of polycrystalline materials,
such as ductile alloys, that are commonly encountered in structures
and machinery components of complex electromechanical systems. The
model of crack damage measure indicates that the fluctuations of
fatigue crack growth are characterized by strong correlation
patterns within short time scales and are uncorrelated for larger
time scales. The two correlation regimes suggest that the 7075-T6
aluminum alloy, analyzed in this paper, is characterized by a
micro-structure which is responsible for an intermittent correlated
dynamics of fatigue crack growth within a certain scale. The
constitutive equations of the damage measure are built upon the
physics of fracture mechanics and are substantiated by
Karhunen-Lo\`{e}ve decomposition of fatigue test data. Statistical
orthogonality of the estimated damage measure and the resulting
estimation error is demonstrated in a Hilbert space setting.
\newline

\end{abstract}


\maketitle

\section{ Introduction}\label{Introduction}

The fracture of solids and the growth of cracks is a typical
instability phenomena which are known to be strongly nonlinear.
Herein we apply to fracture mechanics some of the recent methods
developed in statistical physics. In particular, we use the notion
of fractal statistics to describe the correlation of the
fluctuations around fatigue crack growth in polycrystalline
materials, such as ductile alloys. In this paper, we have
investigated the fatigue fracture properties of 7075-T6 aluminum
alloy.

 The importance of this investigation is that,  in both the physics
and engineering literature, the fluctuations around fatigue crack
growth in a typical material have always been  assumed to be random
or uncorrelated noise. Consequently, the associated models include
uncorrelated random processes. For example, in agrement with the
existent theory of micro-level fatigue cracking, Bogdonoff and Kozin
\cite{Bogdanoff} proposed a Poisson-like uncorrelated-increment jump
model of fatigue crack phenomena.  An alternative approach to
stochastic modeling of fatigue crack damage is to randomize the
coefficients of an existing deterministic model to represent
material inhomogeneity \cite{Ditlevsen}. A third approach has been
to adopt a deterministic model of fatigue crack growth in addition
to a random process, see for example \cite{Lin,Spencer,Ishikawa}.

The fatigue crack growth process can also be modeled by nonlinear
stochastic differential equations using It\^{o} statistics
\cite{Kloeden}  that again presuppose randomness of the
fluctuations. Specifically, the Kolmogorov forward and backward
diffusion equations, which require solutions of nonlinear partial
differential equations, have been proposed to generate the
statistical information required for risk analysis of mechanical
structures \cite{Tsurui,Bolotin}. These nonlinear partial
differential equations have only been solved numerically and the
numerical procedures are computationally intensive as they rely on
fine-mesh models using finite-element or combined finite-difference
and finite-element methods \cite {Sobczyk}. Casciati \textit{et al}.
\cite{Casciati} have analytically approximated the solution of the
It\^{o} equations by Hermite moments to generate a probability
distribution function of the crack length.

Several studies have determined that the stochastic fluctuations
observed in  innumerable natural phenomena   are not   simply
random, that is, uncorrelated noise, but present correlation
patterns that reveal complex and alternative dynamics and/or
material microstructures. Thus,  the purpose of the present research
is to determine  whether uncorrelated stochastic models such as
those previously discussed in the literature are realistic in
describing the fluctuations around fatigue crack growth in
polycrystalline materials, or whether such fluctuations present
patterns that would reveal  complex material micostructure requiring
alternative correlated stochastic models. Two main classes of
correlation patterns are commonly observed in natural time series
and these are denoted as short and long-time correlations.
Short-time  correlations are characterized by phenomena that rapidly
lose memory of past or distant events. This happens, for example,
when the autocorrelation function of the time series decays
exponentially in the time separation between two elements. By
contrast, long-time correlations are characterized by
autocorrelation functions that decay more slowly than (negative)
exponentials; one example is the inverse power-law decay.

 A simple  model, which has been extensively used in the
interpretation of stochastic fluctuations in a time series
 $\{\xi_i\}$ with $i=1,2,\dots,N$, is based on the
evaluation of the mean-square displacement of the  diffusion-like
processes generated by trajectories  $X_n(t)$ defined as
\begin{equation}  \label{eq0}
X_n(t)=\sum_{j=1}^{t} \xi_{n+j}~.
\end{equation}
If $\{\xi_i\}$ is a white random sequence, the diffusion process is
a well-known Brownian motion.  The central limit theorem applied to
the diffusion distribution generated by trajectories $X_n(t)$ yields
a probability density that converges to a Gaussian function whose
mean-square displacement converges asymptotically to
\begin{equation}  \label{eq1}
\left\langle {X(t)^2} \right\rangle
\propto t^\alpha~,
\end{equation}
with $\alpha=1$. In general, it is possible to have  anomalous
behavior yielding  enhanced diffusion ($\alpha >1$) that has been
known for twenty years to arise in dynamically chaotic systems
\cite{Chirikov}, or sublinear diffusive growth ($\alpha <1$) that is
familiar from disordered fractal materials \cite{Havlin}.

 Anomalous diffusion reveals persistent (for an enhanced
diffusive growth) or antipersistent (for a sublinear diffusive
growth) correlation patterns in the dynamics of a random walk. A
persistent random walk is characterized by a probability of stepping
in the direction of the previous step that is greater than that of
reversing directions. An antipersistent random walk is characterized
by a probability of stepping in the direction of the previous step
that is less than that of reversing directions. Sometimes a
momentarily initial enhanced or  sublinear diffusive growth, lasting
up to a certain time-scale, is generated by the  statistical
transition to the asymptotic regime of the diffusion process. For
example, a simple discrete  random walk is described by a binomial
distribution that only asymptotically converges to a Gaussian while
initially presenting an enhanced  diffusive growth \cite{scafetta0}.
Thus, a real autocorrelated time series will lose its correlation
patterns if the temporal order of the sequence is randomized.

 There are a number of different theoretical approaches that
explain the anomalous diffusion depicted in (\ref{eq1}). One such
explanatory model is that of an infinitely long correlated
random walk in which $\alpha =2H$, where $H$ is the Hurst exponent in the interval $%
0\le H\le 1$ with the case $H=0.5$ corresponding to a simple random
walk.  This model has been used extensively in the interpretation of
fluctuations in time series in the physical and life sciences
\cite{West} and is called fractional Gaussian noise \cite
{Mandelbrot}. Another kind of anomalous diffusion has to do with
taking steps that are uncorrelated in time, but on a random or
fractal, not a regular lattice. In the second model, an anomalous
diffusion occurs because geometrical obstacles exist on all length
scales and such obstacles inhibit transport. Havlin and Ben-Avraham
\cite{Havlin} point out that the anomalous exponent $\alpha$ is
related to the fractal dimension of the random walk path on the
lattice. There is a third possible explanation of the anomaly in
(\ref {eq1})
 called a L\'evy walk \cite{Shlesinger} that was first used to understand turbulent
diffusion \cite{Shlesinger} and yields $\alpha \approx 3$, which is
consistent with Richardson's law of enhanced diffusion
\cite{Richardson}.

 Physical examples of  anomalous diffusion processes are
earthquakes \cite{Gutenberg}, rainfall \cite {Shlesinger,Peters},
turbulent fluid flow \cite{Frisch}, relaxation of stress in
viscoelastic materials \cite{West,Glockle}, solar flares \cite
{scafetta1,scafetta2,scafetta3}, and other processes with slip-stick
dynamics.  Recently, a multi-scaling comparative analysis to
distinguish L\'evy walk intermittent noise from fractal Gaussian
intermittent noise was suggested by Scafetta and West
\cite{scafetta4}.

 Finally, a physical system  might be characterized by
different values of the scaling exponent $\alpha$ at different
scales \cite{scanicola1}. Usually, this means that one system is
characterized by a non-self affine structure. The scale at which the
transition from a scaling regime to another occurs indicates the
scale at which the structure changes.  In this work  we determine
that the fluctuations around the ballistic growth of fatigue cracks
in ductile alloys present such a scale transition from a strongly
correlated regime at short-time scales to a random regime at longer
time-scales. Properties, such as grain size distribution, degree of
heterogeneity, the existence of microscopic defects, inclusions,
twin boundaries and dislocations, of polycrystalline materials may
contribute to the micro-mechanisms of fatigue fracture revealed by
the present analysis.

This paper is organized into six sections, including the present
one, and an Appendix. Section \ref{FatigueMeasure} provides the
underlying phenomenology of the stochastic damage measure. Section
\ref{KLdecomposition} presents Karhunen-Lo\`{e}ve (KL) decomposition
of fatigue test data to formulate an estimate of the stochastic
measure, which is statistically orthogonal to the estimation error.
Section \ref{DataAnalysis} focuses on identification of the model
parameters and their probability distributions. Section
\ref{Simulation} presents the results of model prediction by Monte
Carlo simulation. The paper is summarized and concluded in Section
\ref{Conclusions} with recommendations for future research.

\section{ Measure of Fatigue Crack Damage}\label{FatigueMeasure}

 Traditionally fatigue crack growth models have been
formulated by fitting estimated mean values of fatigue crack
length $\hat{a}_t$, generated from ensemble averages of
experimental data, as functions of time in units of cycles \cite
{Paris,Schijve}. Ray and Patankar \cite{Ray2} have formulated the
state-space modeling concept of crack growth based on
fracture-mechanistic principles of the crack-closure concept
\cite{Elber}. The state-space model has been validated by fatigue
test data for variable-amplitude cyclic loading, see for example
Refs. \cite{Schijve,Porter,McMillian}.

 The three panels in Figure 1 show test data of cumulative
fatigue crack growth in the 7075-T6 aluminum alloy under different
cyclic loading~\cite{Ghonem}. It is important to note that the crack
growth curves do not increase smoothly, but they exhibit
fluctuations
 around an ideal smooth curve of
crack growth representing ballistic growth. In this context, a major
objective of the paper is to investigate the autocorrelation
properties of these fluctuations  with the smooth curve removed. In
the following we briefly review the theory and the standard
phenomenological equations that describe the  fatigue crack growth.

In linear fracture mechanics, it is assumed that the stressed
material remains elastic and undamaged everywhere, except in a small
domain in the vicinity of the crack tip. However, this view is not
confirmed by experimental evidence and the process of fatigue damage
accumulation could occur throughout the stressed volume. Paris and
Erdogan \cite{Paris} originally developed a phenomenological model
of crack growth rate, which depends on the stress history and is
thus represented by a continuum rate equation having the hereditary
structure. This model has been subsequently modified by many
researchers (see, for example, citations in Refs.
 \cite{Rabotonov,Anderson,Bannantine}) in the following form.
\begin{equation}\label{basiceq}
\delta \hat{a}_{t}\equiv \hat{a}_{t}-\hat{a}_{t-\delta t} =
h\left(\Delta K^{eff}_{t}\right) \delta t ,
\end{equation}
with $h(0)=0$ and $\hat{a}_{t_{0}}>0$ for $t\geq t_{0}$,
 where $\hat{a}_{t}$ is the estimated mean of the crack length
at the time $t$ during a stress cycle and $\delta t$ is the time
duration of the stress cycle; and $\Delta K^{eff}_{t}$ is the stress
intensity factor range at time $t$, which is given by the
experimentally validated empirical model.
\begin{equation}\label{stressIntensity}
\Delta K^{eff}_{t}=\Delta S_{t} \sqrt{\pi \ \hat{a}_{t-\delta t}} \
F(\hat{a}_{t-\delta t}),
\end{equation}
where $\Delta S_{t}$ is the range (i.e., the difference between
maximum and minimum values) of the stress cycle at time $t$, which
is directly related to the applied load. Experimental observations
suggest that both duration and shape of a stress cycle are not
relevant for crack growth in ductile alloys at room temperature. A
stress cycle is only characterized by the minimum stress $S^{min}$
and the maximum stress $S^{max}$, respectively, and is denoted as
the ordered pair $(S^{min},S^{max})$. The empirical relation
$F(\bullet)$ in Eq. (\ref{stressIntensity}) represents the
geometry of the crack tip; for center-cracked specimens of
half-width $w$ with $0<\hat{a}_{t}<w$ at all $t\ge t_{0}$, the
structure of $F(\bullet)$ has been experimentally determined
as~\cite{Anderson}:
\begin{equation}\label{empirical}
    F(\hat{a}_{t-\delta t})=\sqrt{\sec{\left(\frac{\pi }{2 w} \ \hat{a}_{t-\delta
    t}\right)}}.
\end{equation}

The function $h(\bullet)$ in Eq. (\ref{basiceq}) is a non-negative
Lebesgue-measurable function that is dependent on the material and
geometry of the stressed component. It has been shown in the
fracture mechanics literature \cite{Anderson,Bannantine} that, for
center-cracked specimens of ductile alloys, the function
$h(\bullet)$ obeys the power law:
\begin{equation}\label{powerLaw}
h\left(\Delta K^{eff}_{t}\right) = \left(\Delta
K^{eff}_{t}\right)^{m},
\end{equation}
where the exponent parameter $m$ is dependent on the material of the
stressed component; for ductile alloys, $m$ is in the range of 2.5
to 5.0 \cite{Anderson}.

Equations (\ref{basiceq}), (\ref{stressIntensity}),
(\ref{empirical}) and (\ref{powerLaw}) are now combined to formulate
a mean-value model of fatigue crack growth for center-cracked
specimens of ductile alloy materials:
\begin{equation}\label{meanValue}
\delta \hat{a}_{t} \propto~ \left[\Delta
S_{t}\sqrt{\hat{a}_{t-\delta t} \ \sec{\left(\frac{\pi }{2 w} \
\hat{a}_{t-\delta t}\right)}}\right]^{m}   \delta t .
\end{equation}
with $\hat{a}_{t_{0}}>0 $ and $t\geq t_{0}$.

 Following Sobczyk and Spencer \cite{Sobczyk} and the pertinent
references cited therein, we randomize the deterministic mean-value
model, Eq. (\ref{meanValue}), to obtain a stochastic model for the
rate of crack growth. The stochastic model of continuous crack
length is built upon the model structure proposed by
Ray~\cite{Ray2,Ray1}, and is given by:
\begin{equation}\label{stochastic}
d {c}_{t}(\zeta)={\Omega \left( {\zeta ,t}\right) }\left[ \Delta
S_{t}~\sqrt{\frac{{c}_{t}(\zeta)}{\cos \left(
\frac{\pi }{2}~{c}_{t}(\zeta)\right) }}\right] ^{m}~d t\cong ~{\Omega \left( {\zeta ,t}\right) }\frac{%
\left( \Delta S_{t}~\sqrt{{c}_{t}(\zeta)}\right) ^{m}}{1-m\left( {\frac{\pi }{4}~{c}%
_{t}(\zeta)}\right) ^{2}}~d t,
\end{equation}
where the random sample $\zeta$ signifies a specimen or a machine
component on which  a fatigue test is conducted; the dimensionless
stochastic crack length ${c}_{t}(\zeta)$ is normalized with respect
to the half width $w$,  that is, the mean value $\hat{c}_{t}\equiv
\hat{a}_{t}/w$. Equation (\ref{stochastic}) is a continuous
stochastic version  of Eq. (\ref{meanValue}), where the differential
of the stochastic crack length $d {c}_{t}(\zeta)$ is a function of
the  crack length ${c}_{t}(\zeta)$ at time $t$  and the normalized
stress $\Delta S_{t}\equiv \Delta S_{t}^{e}/S^y$, where $S^y$ is the
yield stress of the material. The condition $0<c_{t_{0}}\leq
c_{t}<\frac{4}{\pi \sqrt{m}}$ is imposed to ensure non-negativity of
the crack length increment almost surely, i.e., $d c_{t}(\zeta)>0$
for almost all samples $\zeta$.  The stochastic process of crack
growth is largely dependent on the second-order random process
$\Omega(\zeta ,t)$ and the exponent parameter $m$ in
Eq.(\ref{stochastic}).

To investigate the stochastic properties of the fatigue crack
growth process, we separate $\Omega(\zeta ,t)$ into two parts as:
\begin{equation} \label{Omega}
\Omega \left(\zeta ,t\right) =\Omega _{0}\left(\zeta \right) \left[1
+ \Omega _{1}\left( {\zeta ,t}\right)\right],
\end{equation}
where the time-independent component $\Omega _{0}\left( \zeta
\right)$ represents uncertainties in manufacturing, for example in
machining, and makes a major contribution to the ballistic component
of the crack growth; the time-dependent component $\Omega _{1}\left(
{\zeta ,t}\right)$ represents uncertainties in the material
microstructure and crack length measurements that may vary with
crack propagation in a sample $\zeta$.  This latter component is
primarily  responsible for the small fluctuations around the
ballistic component of crack growth whose autocorrelation properties
we  study.

We postulate that $\Omega _{0}$ and $\Omega _{1}$ in Eq.
(\ref{Omega}) are statistically independent of one another for all
$t\geq t_{0}$, where $t_0$ is the initial time. The rationale for
this independence assumption is that inhomogeneity of the material
microstructure and measurement noise, associated with each test
specimen and represented by $\Omega _{1}\left( {\zeta ,t}\right) $,
are unaffected by the uncertainty $\Omega _{0}\left( \zeta \right)$
due, for example, to machining operations. Without loss of
generality, we assume that the fluctuations in time have a zero mean
value, i.e., $\left\langle {\Omega _{1}\left( {\zeta ,t}\right) }
\right\rangle =0\mbox{ for all }t\ge t_{0}$. Furthermore,
non-negativity of the crack growth rate $d c_t\left(\zeta\right)$ in
Eq. (\ref{stochastic}) is assured in the almost sure (\textit{a.s.})
sense by imposing the constraint $\Omega _{0}\left(\zeta \right)\ge
0$ with probability 1 (wp 1).

 For notational brevity, let us suppress the term $\zeta$ in
random processes like $c_t(\zeta)$ and $\Omega \left( {\zeta
,t}\right)$. A combination of Eqs. (\ref{stochastic}) and
(\ref{Omega})  and few simple algebraic steps yield  the following
 equation for each sample point $\zeta$:
\begin{equation} \label{Omega1}
\left[ {c_{t}^{-m/2}-m\left( {\frac{\pi }{4}}\right)
^{2}c_{t}^{2-m/2}} \right] d c_{t} = (\Delta S_{t})^{m} \ \Omega
_{0}\left[1 + \Omega _{1}\left(t\right)\right] d t \ \ w.p. 1.
\end{equation}
Pointwise integration of Eq. (\ref{Omega1}) yields the solution of
fatigue damage increment from the initial time $t_o$ to the
current time $t$ as:
\begin{equation}  \label{Omega2}
\psi \left( {t,t_0 } \right) = \int\limits_{t_0 }^t (\Delta
S_{t'})^m \ \Omega _0 \left[1 + \Omega _1 \left(t^{\prime}\right)
\ \right]dt^{\prime} \ \ w.p. 1
\end{equation}
An explicit  expression of the stochastic diffusion process $\psi
\left( {t,t_0 } \right)$ is obtained by integrating the left side
of Eq. (\ref{Omega1}) and is given by
\begin{equation}  \label{Omega5}
\psi \left( {t,t_0 } \right)\equiv \left[ {\frac{c_t^{1-m/2}-c_{t_0 }^{1-m/2}%
}{1-m/2}} \right]-m\left( {\frac{\pi }{4}} \right)^2\left[ {\frac{%
c_t^{3-m/2}-c_{t_0 }^{3-m/2}}{3-m/2}} \right]
\end{equation}
where $\psi \left( {t,t_0 } \right)$ represents a dimensionless
non-negative measure of fatigue crack damage increment from the
initial instant $t_0 $ to the current instant $t$ as a function of
the normalized crack length. The constant parameter $m$ in
(\ref{Omega5}) is in the range of 2.5 to
5 for ductile alloys and metallic materials ensuring that $(1-m/2)<0$ and $%
(3-m/2)>0$. The diffusion process $\psi \left( {t,t_0 } \right)$
is almost surely continuous because it is a continuous function of the crack length process $%
c_t \ \it{wp}$ 1. Both $c_t$ and $\psi \left( {t,t_0 } \right)$
are measurable functions although their (probability) measure
spaces are different. In essence, the probability of $\psi \left(
{t,t_0 } \right)$, conditioned on the initial crack length $c_{t_0
}$, leads to a stochastic measure of fatigue crack damage
increment at the instant $t$ starting from the initial instant
$t_0 $.

For a constant stress range $\Delta S$, we carry out the time
integration in Eq. (\ref{Omega2}) to obtain
\begin{equation}  \label{Omega3}
\psi \left( {t,t_0 } \right) = (\Delta S)^m \left[\Omega _0
(t-t_0) + \Theta (t,t_0)\right]
\end{equation}
where the second term  on the right side is the time integral
\begin{equation}  \label{Omega4}
\Theta \left( {t,t_0 } \right)\equiv \Omega _0 \int\limits_{t_0 }^t {\Omega _1 \left( {%
t^{\prime}} \right)} dt^{\prime}.
\end{equation}
 Thus, the stochastic diffusion process $\psi \left( {t,t_0 }
\right)$ according to the model (\ref{Omega3}) is given as the sum
of a random component, linear in time, plus a time-fluctuating
component proportional to the diffusion process $\Theta \left(
{t,t_0 } \right)$.

The objective is to validate the model in Eq. (\ref{Omega2}) by
decomposing the damage increment measure $\psi \left( {t,t_0 }
\right)$ into two parts that are mutually statistically
independent and, at the same time,  equivalent to the two
components of the right side of Eq. (\ref{Omega3}). That is, we
would like to obtain an estimate $\hat {\psi }\left( {t,t_0 }
\right)$ of the stochastic damage increment measure $\psi \left(
{t,t_0 } \right)$ and of the fluctuations $\tilde {\psi }\left(
{t,t_0 } \right)$ around $\hat {\psi }\left( {t,t_0 } \right)$
from the initial instant $t_0$ to the current instant $t$ such
that:
\begin{equation}  \label{Omega6}
\psi \left( {t,t_0 } \right)\mathop =\limits^{ms} \hat {\psi }\left(
{t,t_0 } \right)+\tilde {\psi }\left( {t,t_0 } \right) ~,
\end{equation}
where $\hat {\psi }\left( {t,t_0 } \right)$ is statistically
equivalent to $\Delta S^m \Omega _0 (t-t_0 )$, and $\tilde {\psi
}\left( {t,t_0 } \right)$ is statistically equivalent to $(\Delta
S)^m \Theta (t,t_0 )$ of Eq. (\ref{Omega3}).

 To test the validity of the above postulate that the two
components of the multiplicative random process $\Omega
_{0}\left(\zeta \right) $ and $ \Omega _{1}\left({\zeta ,t}\right)$
in Eq. (\ref{Omega}) are statistically independent, we require that
the zero-mean estimation error $\tilde {\psi }\left( {t,t_0 }
\right)$ be statistically orthogonal to the estimate of the
increment measure $\hat {\psi }\left( {t,t_0 } \right)$ in the
Hilbert space $L_2 \left(P\right)$ defined by the probability
measure $P$. As such $\hat {\psi }\left( {t,t_0 } \right)$ is the
best linear estimate of the stochastic diffusion process. Based on
mean-square continuity of the damage measure $\psi \left( {t,t_0 }
\right)$, the next section elaborates on the model structure laid
out in Eq. (\ref{Omega6}). To this end, we analyze experimental data
sets of random fatigue via Karhunen-Lo\`{e}ve (KL) decomposition
\cite {Jazwinski,Wong,Fukunaga} that guarantees the above
statistical orthogonality among the components of the decomposition.
In Section IV we also use these experimental data sets to identify
the model parameters.

\section{Karhunen-Lo\`{e}ve Decomposition of Experimental
Data}\label{KLdecomposition}

In this section we analyze fatigue test data via KL-decomposition
\cite {Fukunaga} to justify the model structure postulated in Eqs.
(\ref{Omega2}) and (\ref{Omega5}). We use the experimental data of
random fatigue crack growth in the 7075-T6 aluminum alloy
\cite{Ghonem} and conduct the tests under different constant load
amplitudes at ambient temperature. For all experiments the
half-width is $w = 50.8 mm$, the initial crack length is $a_{t_0}
= 9 mm$, and, therefore, the initial dimensionless crack length is $%
c_{t_0}=a_{t_0}/w=0.18$ with probability 1. The Ghonem data sets
were generated for 60 center-cracked specimens each at three
different constant load amplitudes: (i) Set \#1 with peak nominal
stress of 70.65 MPa (10.25 ksi) and stress ratio $R\equiv
S^{min}/S^{max}$ = 0.6 for 54,000 cycles, the effective stress range
$\Delta S^e$= 15.84 MPa; (ii) Set \#2 with peak nominal stress of
69.00 MPa (10.00 ksi) and $R$ = 0.5 for 42,350 cycles, and $\Delta
S^e$= 17.80 MPa; and (iii) Set \#3 with peak nominal stress of 47.09
MPa (6.83 ksi), $R$ = 0.4 for 73,500 cycles, and $\Delta S^e$= 13.24
MPa.  The three experimental datasets \cite{Ghonem} are shown in the
three panels of Figure 1.

The KL-decomposition requires the mean and covariance of the
stochastic measure of damage increment $\psi \left( {t,t_0 }
\right)$ which are expressed as:
\begin{equation}  \label{eq11}
\begin{array}{l}
\mu _\psi \left( {t,t_0 } \right)\equiv \left\langle {\psi \left(
{t,t_0 }
\right)} \right\rangle \\
\\
C_{\psi \psi } \left( {t_1 ,t_2 ;t_0 } \right)\equiv \left\langle
[{\psi \left( {t_1 ,t_0 } \right)-\mu _\psi \left( {t_1,t_0 }
\right)][\psi \left( {t_2 ,t_0 } \right)-\mu _\psi \left( {t_2,t_0 }
\right)]} \right\rangle
\end{array}
\end{equation}
The covariance function $C_{\psi \psi } (t_1 ,t_2 ;\,t_o )$ in Eq.
(\ref {eq11}) is continuous at $t_1 =t_2 =t$ for all $t\ge t_0 $.
Hence, the process $\psi \left( {t,t_0 } \right)$ is mean-square
(\textit{ms}) continuous based on a standard theorem of mean-square
calculus \cite {Jazwinski,Wong}.  The mean and covariance are
calculate for the 60 available center-cracked specimens in each
case.

Since only finitely many data points at $n$ discrete instants are available
from experiments, an obvious approach to the analysis of the damage estimate
is to discretize over the finite time horizons $\left[ t_0,t \right]$ so
that the stochastic process $\psi \left( {t,t_0 } \right)$ becomes the $n$%
-dimensional random vector $\mathrm{\mathbf{\psi }}$. Consequently, the
covariance function $C_{\psi \psi } (t_1 ,t_2 ;t_o )$ in Eq. (\ref{eq11}) is
reduced to a real semipositive-definite $(n\times n)$ symmetric matrix $%
\mathrm{\mathbf{C}}_{\psi \psi } $. Since the experimental data were
collected at sufficiently close intervals,
$\mathrm{\mathbf{C}}_{\psi \psi } $ contains pertinent information
of the crack damage process. The $n$ (real non-negative) eigenvalues
of $\mathrm{\mathbf{C}}_{\psi \psi } $ are ordered as $\lambda _1
\ge \lambda _2 \ge \,\cdots \,\ge \lambda _n $, with the
corresponding eigenvectors, $\varphi ^1,\,\,\varphi ^2,\,\,\cdots
\,\,,\,\,\varphi ^n$, that form an orthogonal basis of $\Re ^n$ for
signal decomposition. The KL-decomposition also ensures that the $n$
random coefficients of the basis vectors are statistically
orthogonal, that is, they have zero mean and are mutually
uncorrelated. These random coefficients form a random vector
$\mathrm{\mathbf{X}}\equiv \left[ {x_1 \;,x_2
,\,\;\cdots \;\,,x_n } \right]^T$ having the covariance matrix $\mathrm{%
\mathbf{C}}_{XX} =diag\;\left( {\lambda _1 ,\,\lambda _2 ,\,\,\cdots
\,,\,\lambda _n } \right)$ leading to a decomposition of the discretized
signal as:
\begin{equation}  \label{eq15}
\mathrm{\mathbf{\psi }}\mathop =\limits^{ms} \left\langle \mathrm{\mathbf{%
\psi }} \right\rangle +\sum\limits_{j=1}^n {x_j } \phi ^j
\end{equation}
It was observed by Ray~\cite{Ray1} that the statistics of crack
length are dominated by the random coefficient corresponding to
the principal eigenvector (i.e., the eigenvector associated with
the largest eigenvalue) and that the combined effects of the
remaining eigenvectors are small. Therefore, the signal
$\mathrm{\mathbf{\psi }}$ in Eq. (\ref{eq15}) is expressed as the
sum of a principal part and a (zero-mean) residual part that are
mutually statistically orthogonal:
\begin{equation}  \label{eq16}
\mathrm{\mathbf{\psi }}\mathop =\limits^{ms} \mbox{ }\mathop {\left\langle
\mathrm{\mathbf{\psi }} \right\rangle +x_1 \phi ^1}\limits_{%
\mbox{principal
part}} \mbox{ }+\mbox{ }\mathop {\sum\limits_{j=2}^l {x_j } \phi ^j}\limits_{%
\mbox{residual part}}
\end{equation}
Thus, as Eq. (\ref{Omega6}) requires,   the  vector
$\mathrm{\mathbf{\psi }}$ is expressed as the sum of the principal
and residual parts with equality in the mean square (\textit{ms})
as:
\begin{equation}  \label{eq17}
\mathrm{\mathbf{\psi }}\mbox{ }\mathop =\limits^{ms} \mbox{ }\mathrm{\mathbf{%
\hat {\psi }}}\mbox{ }+\mbox{ }\mathrm{\mathbf{\tilde {\psi }}}
\end{equation}
where the principal part is the damage estimate
\begin{equation}  \label{eq1uu8}
\mathrm{\mathbf{\hat {\psi }}}~~\equiv ~~\left\langle \mathrm{\mathbf{\psi }}
\right\rangle \mbox{ }+\mbox{ }x_1 \phi ^1~,
\end{equation}
the residual part is the estimation error representing the
fluctuations around the mean damage estimate (\ref{eq1uu8})
\begin{equation}  \label{eq1tt8}
\mathrm{\mathbf{\tilde {\psi }}}~~\equiv ~~\sum\limits_{j=2}^n {x_j \phi ^j}%
~,
\end{equation}
and the resulting (normalized) mean square error \cite{Fukunaga} is:
\begin{equation}  \label{eq18}
\varepsilon _{rms}^2 \equiv \frac{Trace\left\{ {Cov\left[ {\mathrm{\mathbf{%
\psi }}-\mathrm{\mathbf{\hat {\psi }}}} \right]} \right\}}{Trace\left\{ {%
Cov\left[ \mathrm{\mathbf{\psi }} \right]} \right\}}=\frac{%
\sum\limits_{j=2}^n {\lambda _j } }{\sum\limits_{j=1}^n {\lambda _j } }.
\end{equation}
The KL-decomposition of fatigue test data sets reveals that $0.01\le
\varepsilon _{rms}^2 \le 0.1$ for all three data sets.

The principal eigenvector $\phi ^{1}\left( t\right) $, associated with the
largest eigenvalue $\lambda _{1}$, closely fits the ramp function $\left( {%
t-t_{0}}\right) $ for each of the three data sets in Figure 1;
this is shown in Figure 2 for the data set 1. Comparing the terms
on the right hand side of the discrete model in Eq. (\ref {eq17})
with those of the continuous model in Eq. (\ref{Omega3}), it is
reasonable to have the random variable $\Delta S^{m}\left[ {\Omega
_{0}-\mu _{0}}\right] $ equal (in \textit{ms} sense) to the random
coefficient $x_{1}$ of the principal eigenvector $\varphi
^{1}(t)$. Applying the lemma from the Appendix, a mean-square
equivalence between the KL-decomposition model in
Eq. (\ref{eq17}) derived from the test data and the postulated model in Eq. (%
\ref{eq15}) is established as:
\begin{equation}
\mathop {\langle \psi (t)\rangle }\limits_{\mbox{discrete model (test data)}}%
\mathop \approx \limits^{\mbox{ms}}\mbox{ }\mathop {\Delta S^{m}~{\mu _{0}}%
~\left( {t-t_{0}}\right) }\limits_{%
\mbox{continuous model (constitutive
relation)}}  \label{eq1aa99}
\end{equation}
\begin{equation}
\mathop {x_{1}\phi ^{1}\left( t\right) }\limits_{%
\mbox{discrete model (test
data)}}\mathop \approx \limits^{\mbox{ms}}\mbox{ }\mathop {\Delta
S^{m}\left[ {\Omega _{0}-\mu _{0}}\right] \left( {t-t_{0}}\right) }\limits_{%
\mbox{continuous model (constitutive relation)}}~.  \label{eq1aa9}
\end{equation}
\begin{equation}
\mathop {\sum\limits_{j=2}^{n}{x_{j}}\phi ^{j}}\limits_{%
\mbox{discrete model
(test data)}}\mbox{ }\mathop \approx \limits^{\mbox{ms}}\mbox{ }\mathop {%
(\Delta S)^{m}\Theta \left( {t,t_{0}}\right) }\limits_{%
\mbox{continuous model
(constitutive relation)}}  \label{eq19}
\end{equation}

Thus, we have $\hat {\psi }=\langle \psi (t)\rangle + x_{1}\phi
^{1}\left( t\right) \approx \Delta S^{m} \Omega _{0}\left(
{t-t_{0}}\right)$, and $\tilde {\psi
}=\sum\limits_{j=2}^{n}{x_{j}}\phi ^{j}\approx \Delta S^{m}\Theta
\left( {t,t_{0}}\right)$ as assumed in Eq. (\ref{Omega6}). The two
entities on left hand side in Eqs. (\ref{eq1aa9}) and (\ref
{eq19}) are mutually statistically orthogonal by construction.
Similarly, in view of Eq. (\ref{Omega6}), the zero-mean estimation
error $\tilde{\psi}\left(
{t,t_{0}}\right) $ is statistically orthogonal to $\hat{\psi}\left( {t,t_{0}}%
\right) $ in the Hilbert space $L_{2}\left( P\right) $ defined by the
probability measure $P$ associated with the stochastic process $\psi \left( {%
t,t_{0}}\right) $. As such $\hat{\psi}\left( {t,t_{0}}\right) $ can
be viewed as the best linear estimate of $\psi \left(
{t,t_{0}}\right) $ with the least error $\tilde{\psi}\left(
{t,t_{0}}\right) $ in the mean-square sense.

 It follows from Eqs. (\ref{Omega6}) to (\ref{eq19}) that the
uncertainties associated with an individual sample resulting from
the damage measure estimate $\hat{\psi}\left( {t,t_{0}}\right) $
dominate the cumulative effects of material inhomogeneity and
measurement noise in the estimation error $\tilde{\psi}\left(
{t,t_{0}}\right) $ unless $\left( {t-t_{0}}\right) $ is small.
Therefore, from the perspectives of material-health monitoring, risk
analysis, and remaining life prediction where the inter-maintenance
interval $\left( {t-t_{0}}\right) $ is expected to be large, a
reasonably accurate identification of the mean $\mu _{0}$ and
variance $\sigma _{0}^{2}$ of the random parameter $\Omega _{0}$ is
crucial, while the role of the diffusion process $\Theta \left(
{t,t_{0}}\right) $ is relatively less significant. This observation
is consistent with the statistical analysis of fatigue test data by
Ditlevsen \cite{Ditlevsen} where the random process described by Eq.
(\ref{eq19}) is treated as the zero-mean residual. Ditlevsen
\cite{Ditlevsen} also observed largely similar properties by
statistical analysis. Nevertheless, the stochastic properties of
fluctuating function $\Theta \left( {t,t_{0}}\right)$, which we
 investigate, can disclose important information about
the material structure of alloys during crack damage.

\section{ Data analysis} \label{DataAnalysis}

In this section we investigate the stochastic equivalence made in
Eq. (\ref{eq19}) between the residual component of the signal as
obtained by the KL-decomposition and the linear approximation. The
first step is to evaluate the exponent parameter $m$ by fitting the
data of the crack growth with Eq. (\ref{stochastic}). The fit is
done by considering the crack increments from all 60 cases for each
of the three experiments.

By using the empirical values of $m$ it is possible to estimate
$\psi (t,t_{0})$ via Eq. (\ref{Omega5}). The three plots in Figure
2 compare the curve $\psi (t,t_{0})$,  its principal part
according to the KL-decomposition and its linear approximation
according to the continuous model made in Eqs. (\ref{eq1aa99})
plus (\ref{eq1aa9}) for set \#1: the figures for the other data
sets look qualitatively similar. Figure 3 shows the quality of the
equivalence made in Eqs. (\ref{eq1aa99}) plus (\ref{eq1aa9})
between the discrete model, which makes use of the
KL-decomposition, and the continuous model, which makes use of a
linear approximation.

Figure 3 shows the fitted data and the results for  set \#1; the
figures for the other sets are similar. The parameters for all
three sets are listed below.
\begin{itemize}
\item $\Omega \left( {\zeta }\right) ~\Delta S^{m}=0.0019\pm
0.0002$ and $m=4.0\pm 0.2$ for set \#1; \item $\Omega \left(
{\zeta }\right) ~\Delta S^{m}=0.0022\pm 0.0002$ and $m=3.8\pm 0.2$
for set \#2; \item $\Omega \left( {\zeta }\right) ~\Delta
S^{m}=0.0018\pm 0.0002$ and $m=4.7\pm 0.2$for  set \#3.
\end{itemize}

\subsection{Diffusion standard deviation analysis of the fluctuations}

We evaluate the stochastic equivalence made in Eq. (\ref{eq19})
between the residual part of the discrete model, which makes use of
the KL-decomposition, and the residual part of the continuous model,
which makes use of a linear approximation, in two steps. Step 1
compares the size of the increments of the correspondent residual
parts; and Step 2 adopts the standard deviation analysis (SDA) which
is a statistical formalism to study the long-time correlation in a
fractal time series.

Because $\Theta (t)=$ $residual$ $part$, the increments are given by $%
\theta _{t}=\Theta (t)-\Theta (t-1)$. We calculate the standard
deviation, $\sigma _{\theta }$ of the increments $\{\theta _{t}\}$
for each residual component estimated by means of the
KL-decomposition and of the linear approximation respectively.
Finally we calculate the average of the standard deviation, $%
\langle \sigma _{\theta }\rangle $, between the sixty $\sigma
_{\theta }$ for each of the three cases. The results shown in Table
I demonstrate the compatibility of the increments obtained with the
residual parts of the KL-decomposition and the residual part of the
continuous model.

Now, let us suppose that a generic residual curve is given by the function $%
\Theta (t)$, see Eq. (\ref{eq19}), that in this specific case is a
kind of random walk around the ballistic part of the signal, which
is the principal component of the KL-decomposition or the linear
component of the continuous model. The SDA determines the scaling
of the standard deviation of the diffusion process defined as
\begin{equation}
D(\tau )=\frac{1}{\sigma _{\theta }}\sqrt{\sum_{t=0}^{N-\tau }\frac{\left[ \Theta (t+\tau )-\Theta (t)-%
\overline{\Theta (t+\tau )-\Theta (t)}\right] ^{2}}{N-\tau -1}},
\label{sda}
\end{equation}
 where
\begin{equation}
\overline{\Theta (t+\tau )-\Theta (t)}=\sum_{t=0}^{N-\tau
}\frac{\Theta (t+\tau )-\Theta (t)}{N-\tau } ~, \label{sda}
\end{equation}
 $N$ is the number of data points and the times $t$ and $\tau$ are
  measured in cycle period units. It is easy to realize that Eq. (\ref{sda}) ensures that  $D(\tau=1 )=1$.
  In the presence of fractal
statistics we   would have, based on the discussion of anomalous
diffusion in the Introduction,
\begin{equation}\label{}
   D\left( \tau \right) \propto \tau ^{\beta }=\tau ^{\alpha/2 } ~.
\end{equation}
 Figure 4 shows the SDA for the residual part of the
KL-decomposition. Each set of graphs concerning the same crack data
look quite similar. All three sets of graphs show that the curves
have a initial scaling exponent approximately within the range
$0.5<\beta<0.9$.   The mean curve value is represented by the curves
with black circles in Figure 4. These early time values of $\beta$,
interpreted in terms of the random walks discussed in the
Introduction, indicate that the residual parts of the signal
manifest a persistent behavior, that is, a persistent correlation
that lasts at least 10 consecutive cycles on average.

For $10<\tau <100$ the data presents a slight antipersistency with
$0.4<H<0.5$. Consequently, the residual process is initially
strongly persistent, but asymptotically it is almost random. We
observe that for $10<\tau <100$ the mean scaling exponent is
approximately $H=0.45$ in the case of the linear continuous model
and this is slightly larger than the scaling exponent in the KL
discrete decomposition. This change in scaling is due to the fact
that the principal part obtained with the KL decomposition
extracts more information from the original signal than does the
simple linear approximation.

In the introduction we have explained that  an initial anomalous
diffusion that last up to a certain $\tau$ as detected by Eq.
(\ref{eq0}) could also be an artifact   related not to some
autocorrelation pattern in the data but to the transition from the
initial geometrical properties of the distribution of the events
$\{\xi_i\}$ of a time series to the  Gaussian shape of the
asymptotic diffusion distribution. To check that the persistent
behavior for $\tau<10$ observed in the plots of Figure 4 expresses
real correlation patterns, we repeat SDA of the data after
randomizing the time series of the increments $\{\theta_t\}$. That
is, for each crack data first we have the sequence $\{\theta_t\}$
defined as $ \theta _{t}=\Theta (t)-\Theta (t-1)$, then we shuffle
$\{\theta_t\}$ and obtain a new sequence $\{\theta_t'\}$ and
generate a new walk $\Theta' (t)=\sum_{i=1}^t \theta_i'$, and
finally we apply SDA to the new curve $\Theta' (t)$. Figure 5 shows
the result for the crack set \#1 where the residual part is
estimated with the KL-decomposition; for the other datasets the
results are similar. Figure 5 clearly shows that after shuffling of
the temporal order of the single increments $\{\theta_t\}$, the SDA
of the new sequence gives a scaling value of approximately $H=0.5$
and the persistent behavior for $\tau<10$ observed in Figures 4 is
absent. Thus, we conclude that the persistent behavior for $\tau<10$
observed in Figures 4 expresses real correlation patterns in the
fluctuations of crack growth.

Figure 6 also shows that the the distributions of the scaling
exponent seems to be quite uniform in the interval $0.5<\beta <0.9$
(with a probability $P>0.9\%$) or, perhaps, as Figure 6c shows
better, there might be a slight prominence or skewness in favor of
small value of $\beta $. In any case, all figures show that the
distribution of the scaling exponent for the residual components of
the curve obtained with the KL-decomposition or the linear component
of the continuous model practically coincide for all three datasets.
This equivalence suggests that the continuous linear model
essentially captures not only the dominant
properties of the signal, as obtained through the KL-decomposition, see Eq. (%
\ref{eq1aa9}), but also the stochastic properties of the residual signal, as
suggested in Eq. (\ref{eq19}).

\subsection{Statistics of damage measure estimates}

We investigate the statistics of the damage measure estimates
using a lognormal distribution. This is in keeping with the
analysis of several investigators who assumed the crack growth
rate in ductile alloys is lognormal-distributed, see for example,
the citations in Sobczyk and Spencer \cite{Sobczyk}. Other
investigators have treated the crack length as being
lognormal-distributed~\cite{Ray1}, rather than the residual
fluctuations. The results of KL-decomposition in Eqs. (\ref{eq11})
to (\ref{eq17}) are consistent with these assumptions because
$\Omega _{0}$, which dominates the random behavior of fatigue
crack growth, can be considered to be a perfectly correlated
(ballistic) random process, whereas the non-negative,
multiplicative uncertainty term $\Theta \left( {t,t_{0}}\right) $
is a weakly (positively) correlated random process. Yang and
Manning [39] have presented an empirical second-order
approximation to crack growth by postulating a lognormal
distribution of a parameter that does not bear any
physical relationship to $\Delta S$ but is, to some extent, similar to $%
\Omega _{0}\left( {\Delta S}\right) $ in the present model.

Figure 7 shows the histogram of the slopes $\Delta S^{m}~\Omega
_{0}$ of the curves according to the continuous model for the
experimental data presented by Eq. (\ref{eq19}), such as those shown
in Figure 3c. The histograms are fitted with the lognormal
distribution $p(x,\mu ,\sigma )$:
\begin{equation}
p\left( x;\mu ,\sigma \right) =\frac{1}{x\sqrt{2\pi \sigma ^{2}}}~\exp
\left[ -\frac{\left( \ln (x)-\mu \right) ^{2}}{2\sigma ^{2}}\right] ~.
\label{lognor}
\end{equation}
The measured parameters $\mu $ and $\sigma $ are recorded in Table
II. Finally, the parameters $\mu $ and $\sigma $ are function of
$\mu _{0}=\langle x\rangle $ and $\sigma _{0}^{2}=\langle (x-\mu_0
)^{2}\rangle $ as follows:
\begin{equation}
\mu \equiv \ln (\mu _{0})-\sigma ^{2}/2  \label{eq2599}
\end{equation}
and
\begin{equation}
\sigma ^{2}\equiv \ln \left[ {1+\left( {\frac{\sigma _{0}}{\mu _{0}}}\right)
^{2}}\right] .  \label{eq2488}
\end{equation}
Since the random parameter $\Delta S^{m} ~\Omega _{0}$ is not
explicitly dependent on time, its expected value is obtained from
Eq. (\ref{Omega3}) as:
\begin{equation}
\mu _{0}=\langle \Delta S^{m} ~\Omega_0 \rangle =\left\langle\frac{ {\psi \left( {t,t_{0}}\right) } }{ {t-t_{0}} }\right\rangle~,
  \label{mean}
\end{equation}
which is readily determined from the ensemble average estimate from
each of the data sets. Asymptotically in time we find for the
variance of $\Delta S^{m} ~\Omega _{0}$
\begin{equation}
\sigma _{0}^{2}=\langle (\Delta S^{m} ~\Omega_0 - \mu_0)^2 \rangle =\left\langle\left[\frac{ {\psi \left( {t,t_{0}}\right) } }{ {t-t_{0}} } \right]^2\right\rangle  - \mu_0^2~,
\label{variance}
\end{equation}
so that the variance can be determined directly from the ensemble average
estimate from each of the data sets.

\section{Crack Model Simulation}\label{Simulation}

This section presents the results of Monte Carlo simulation of the
fatigue crack damage process based on the model as it emerges from
the stochastic analysis made in the previous section. The model
that we introduce  approximately reproduces the stochastic
properties of both the ballistic or principal part of the fatigue
crack growth and the associated fluctuations around it. The model
consists in generating independently  the fluctuation and the
principal part of the fatigue crack damage in such a way they are
statistically equivalent to the correspondent observations and
then combining them. The crack model simulation is based on four
steps:

\begin{itemize}
\item \emph{Principal part} or ballistic growth: We generate 60
values $\Delta S^{m}\Omega _{0}$, lognormally distributed
according to Eq. (\ref{lognor}) where the parameters $\mu $ and
$\sigma $ are given by the actual fit of the phenomenological
distribution shown in Figure 6 and recorded in Table II. A sample of the curves  $%
\Delta S^{m}\Omega _{0}(t-t_0)$ simulating the dataset \#1 is shown
in Figure 8b.
 \item \emph{Residual part} or fluctuations around the ballistic growth:
We generate 60 fractal gaussian noise sequences $\{\theta
_{t}^{\prime }\}$ each of length $N$ of the original time sequence
and with scaling exponent uniformly distributed in the interval
$0.5<\beta <0.9$. The standard deviation of each sequence is set
equal to the mean standard deviation of the increments of the
residual component of the data reported in Table I. To simulate the
change of scaling exponent from persistent (for $\tau <10$) to almost random (for $%
\tau >10$), we section each fractal time series $\{\theta
_{t}^{\prime }\}$ into segments of length 10 within which the data
would conserve the correlation, and finally we shuffle the position
of these segments in the time series to reproduce a new time series
$\{\theta _{t}\}$. These new time series will have persistent
correlation for $\tau <10$ and uncorrelated randomness for $\tau
>10$. Finally, the curve $\Theta (t)$ is obtained by
integrating the new sequence $\{\theta _{t}\}$ and by detrending
from it its linear component  because the curve $\Theta (t)$ is
supposed having a zero mean. The SDA sample data analysis of an
example of these synthetic residual data simulating the data set \#1
is shown in Figure 8a.

\item The ballistic growth estimated in the principal part and the
associated fluctuations the residual part are combined according
to Eq. (\ref{eq16}) to obtain a simulated damage increment measure
$\psi (t,t_{0})$ for all sixty sequences and for the three
datasets. Figure 8c shows the simulated damage increment measure
$\psi (t,t_{0})$ simulating the dataset \#1.

\item Finally, by using the respective value of the exponent $m$,
reported in Section \ref{DataAnalysis}, for the  dataset \#1  and
an one-dimensional root-finding computer algorithm,
Eq. (\ref{Omega5}) is inverted to obtain a simulated normalized crack length growth curves $%
c_{t}$, as seen in Figure 8d. The similitude between Figure 8d and
Figure 1a is noteworthy and the figures for the other data sets look
qualitatively very similar; hence they are not presented in this
paper.
\end{itemize}

\section{ Summary and Conclusions}\label{Conclusions}

This paper presents a stochastic measure  of fatigue crack damage.
We have focused on the correlation properties of the fluctuations
around fatigue crack growth in ductile alloys.  The model of crack
damage measure indicates that the fluctuations around fatigue crack
growth present strong correlation patterns within short time scales
and are uncorrelated for larger time scales.  These findings suggest
that the random stochastic models adopted in the present literature
for describing the crack growth dynamics should be augmented with
short-time correlated stochastic models.

The damage measure is modeled as an anomalous diffusion process that
is obtained as a continuous function of the current crack length and
of the initial crack length. Perhaps, the randomness in the damage
measure estimate accrues primarily from manufacturing uncertainties
such as defects generated during machining operations because such
macro-defects are expected to drive the ballistic growth of cracks.
This randomness is captured by a single lognormal-distributed random
variable. Instead, the resulting diffusion process of estimated
fluctuations around the ballistic growth of fatigue cracks is
probably due to the inhomogeneity in the structural material because
it is primarily associated with the micro-structure of the material,
and is represented by a non-stationary fractional Brownian motion
model. This non-stationarity manifests itself in the two scaling
exponents occurring at different  scales. Specifically, we observe a
clear transition in the standard deviation analysis from an early
time slope representing a strong persistence, $\beta \approx 0.7$
lasting for approximately $\tau\approx 10$ to a different slope
asymptotically in time representing randomness, $\beta \approx 0.5$.
This transition occurring at $\tau \approx 10$ from a scaling regime
to another indicates the scale at which a structure change of the
ductile alloys occurs.

The constitutive equation of the damage measure is based on the
physics of fracture mechanics and is validated by KL-decomposition
of fatigue test data for 7075-T6 aluminum alloys at different levels
of (constant-amplitude) cyclic load. The damage estimate is
statistically orthogonal to the resulting zero-mean estimation error
in the Hilbert space $L_{2}\left( P\right) $ defined by the
probability measure of the stochastic damage measure. As such, the
damage estimate is often viewed as a best least-square linear
estimate. However, we find that the KL-decomposition is
statistically equivalent to the linear approximation in the
continuum model that can be then used to simulate the fatigue crack
growth in ductile
alloys.\\

\textbf{Acknowledgments}

The authors are grateful to  Professor H. Ghonem of University of
Rhode Island for providing the test data of random fatigue crack
growth. The work reported in this paper has been supported in part
by the Army Research Office under Grant No. DAAD190110640. The first
author thanks the Army Research Office for the support under grant
DAAG5598D0002.\\

\textbf{Appendix: A Supporting Lemma}

Lemma: Let $A\left( \zeta \right)$ and $B\left( \zeta \right)$ be
second-order real random variables; $x\left( {\zeta ,t} \right)$ and $%
y\left( {\zeta ,t} \right)$ be zero-mean mean-square continuous
(possibly non-separable) real random processes; and the real
$g\left( t \right)$ be
almost everywhere continuous on an interval $\Delta $ such that, for all $%
t\in \Delta $, the following conditions hold:

(i) $A\left( \zeta \right)\mathop =\limits^{ms} B\left( \zeta \right)$;

(ii) $\left\langle {A\left( \zeta \right)x\left( {\zeta ,t} \right)}
\right\rangle =0\mbox{ and }\left\langle {B\left( \zeta \right)y\left( {%
\zeta ,t} \right)} \right\rangle =0$.

Then, the following mean-square identity
\[
A\left( \zeta \right)g\left( t \right)+x\left( {\zeta ,t} \right)\mathop
=\limits^{ms} B\left( \zeta \right)g\left( t \right)+y\left( {\zeta ,t}
\right)
\]
yields
\[
\left. {\
\begin{array}{l}
x\left( {\zeta ,t} \right)=y\left( {\zeta ,t} \right) \\
\left\langle {A\left( \zeta \right)y\left( {\zeta ,t} \right)} \right\rangle
=0 \\
\left\langle {B\left( \zeta \right)x\left( {\zeta ,t} \right)} \right\rangle
=0
\end{array}
} \right\} \mbox{ }\forall ~t \in \Delta
\]
Proof: It follows from the above mean-square identity that
\[
Var\left[ {\left\{ {A\left( \zeta \right)-B\left( \zeta \right)}
\right\}g\left( t \right)+\left\{ {x\left( {\zeta ,t} \right)-y\left( {\zeta
,t} \right)} \right\}} \right]=0
\]
which may be expanded to yield:
\[
\begin{array}{l}
Var\left[ {A\left( \zeta \right)-B\left( \zeta \right)} \right]g\left( t
\right)^2+Var\left[ {x\left( {\zeta ,t} \right)-y\left( {\zeta ,t} \right)}
\right] \\
+\left\langle {\left\{ {A\left( \zeta \right)-B\left( \zeta \right)}
\right\}\left\{ {x\left( {\zeta ,t} \right)-y\left( {\zeta ,t} \right)}
\right\}} \right\rangle g\left( t \right)=0
\end{array}
\]
A combination of Condition (i) and Schwarz inequality yields:
\[
Var\left[ {x\left( {\zeta ,t} \right)-y\left( {\zeta ,t} \right)} \right]=0
\]
and the remaining two identities follow from Condition (ii).

\newpage

\begin{table}[tbp]\label{Table1}
\caption{Values of the fiting parameters $\mu$ and $\sigma$ of the
lognormal distribution (\ref{lognor}) of the histograms shown in
Figures 7.}
\begin{tabular}{|c|c|c|}
\hline & $\mu$ & $\sigma$ \\ \hline Set \#1 & 0.58$\pm$0.05 &
0.20$\pm$0.02 \\ \hline Set \#2 & 0.74$\pm$0.05 & 0.16$\pm$0.02 \\
\hline Set \#3 & 0.42$\pm$0.05 & 0.45$\pm$0.04 \\ \hline
\end{tabular}
\end{table}

\begin{table}[tbp] \label{Table2}
\caption{Mean standard deviation of the increments of the residual
part obtained with the K-L decomposition and the residual part of
the continuous model.}
\begin{tabular}{|c|c|c|}
\hline & K-L & Linear Model \\ \hline
Set \#1 : $\langle\sigma_{\theta}\rangle =$ & 0.0024$\pm$0.001 & 0.0025$\pm$%
0.001 \\ \hline
Set \#2 : $\langle\sigma_{\theta}\rangle =$ & 0.0024$\pm$0.001 & 0.0025$\pm$%
0.001 \\ \hline
Set \#3 : $\langle\sigma_{\theta}\rangle =$ & 0.0038$\pm$0.003 & 0.0043$\pm$%
0.003 \\ \hline
\end{tabular}
\end{table}

\newpage
\begin{figure}[tbp]
\epsfig{file=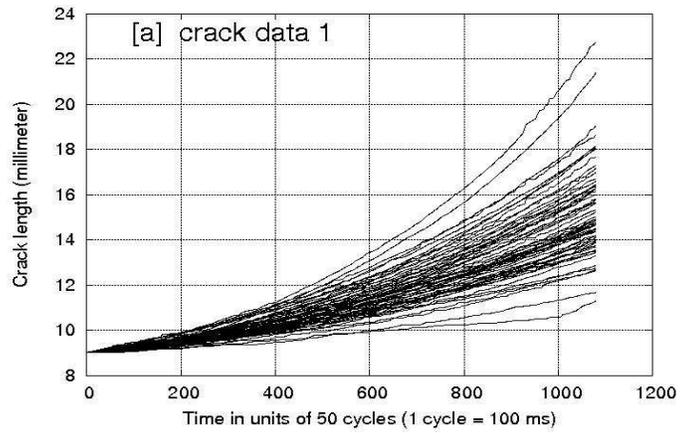,height=10cm,width=7.0cm,angle=-90}
\epsfig{file=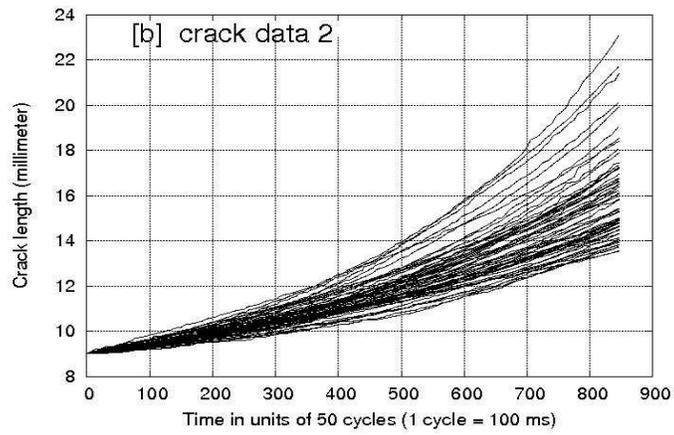,height=10cm,width=7.0cm,angle=-90}
\epsfig{file=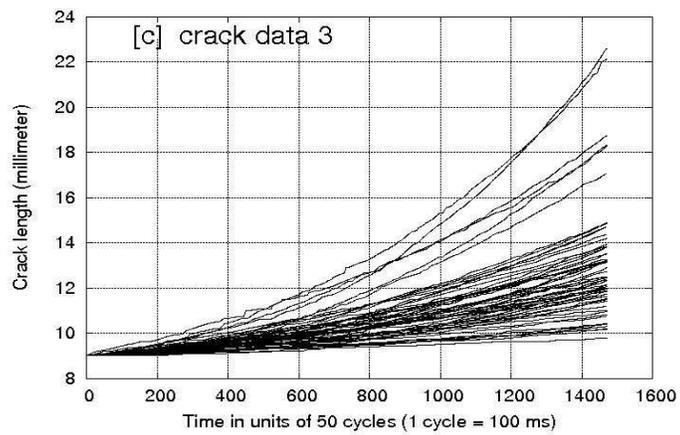,height=10cm,width=7.0cm,angle=-90} \caption{
Experimental data of 7075-T6 aluminum alloy. [a] R=0.6 and Max
stress=70.65 MPa; [b] R=0.6 and Max stress=69.00 MPa; [c] R=0.6 and
Max stress=47.09 MPa. }
\end{figure}

\newpage
\begin{figure}[tbp]
\epsfig{file=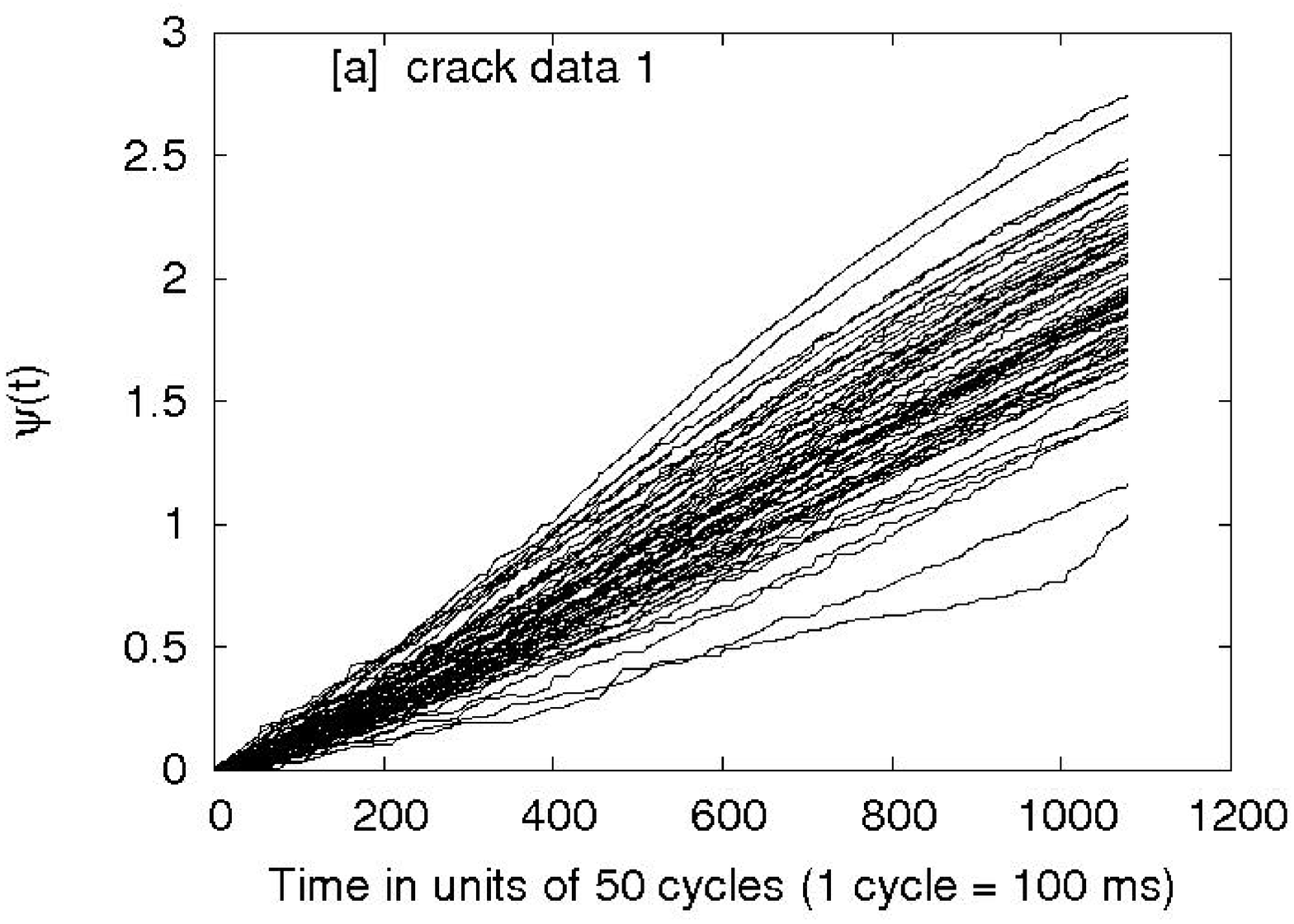,height=10cm,width=7.0cm,angle=-90} %
\epsfig{file=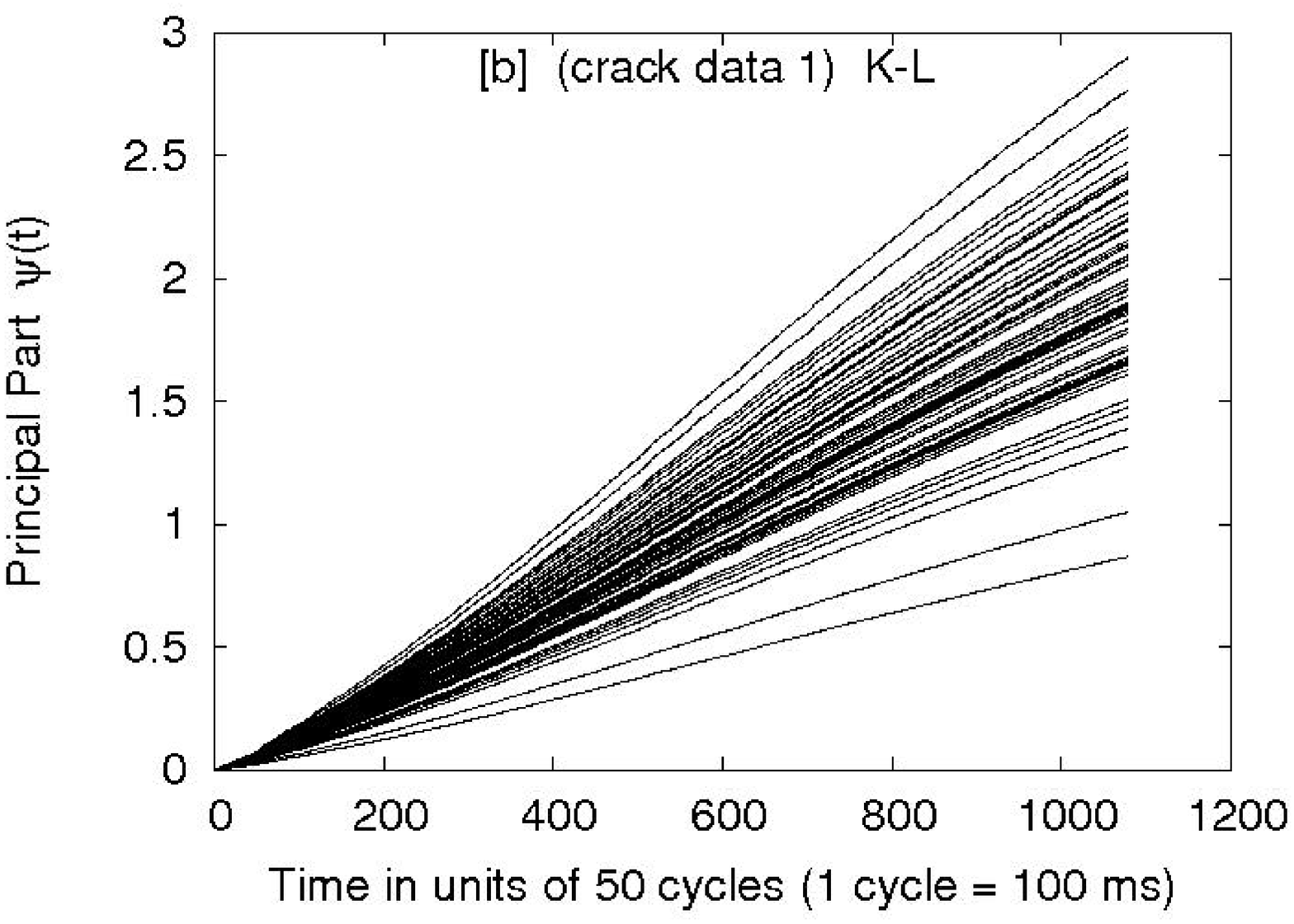,height=10cm,width=7.0cm,angle=-90} %
\epsfig{file=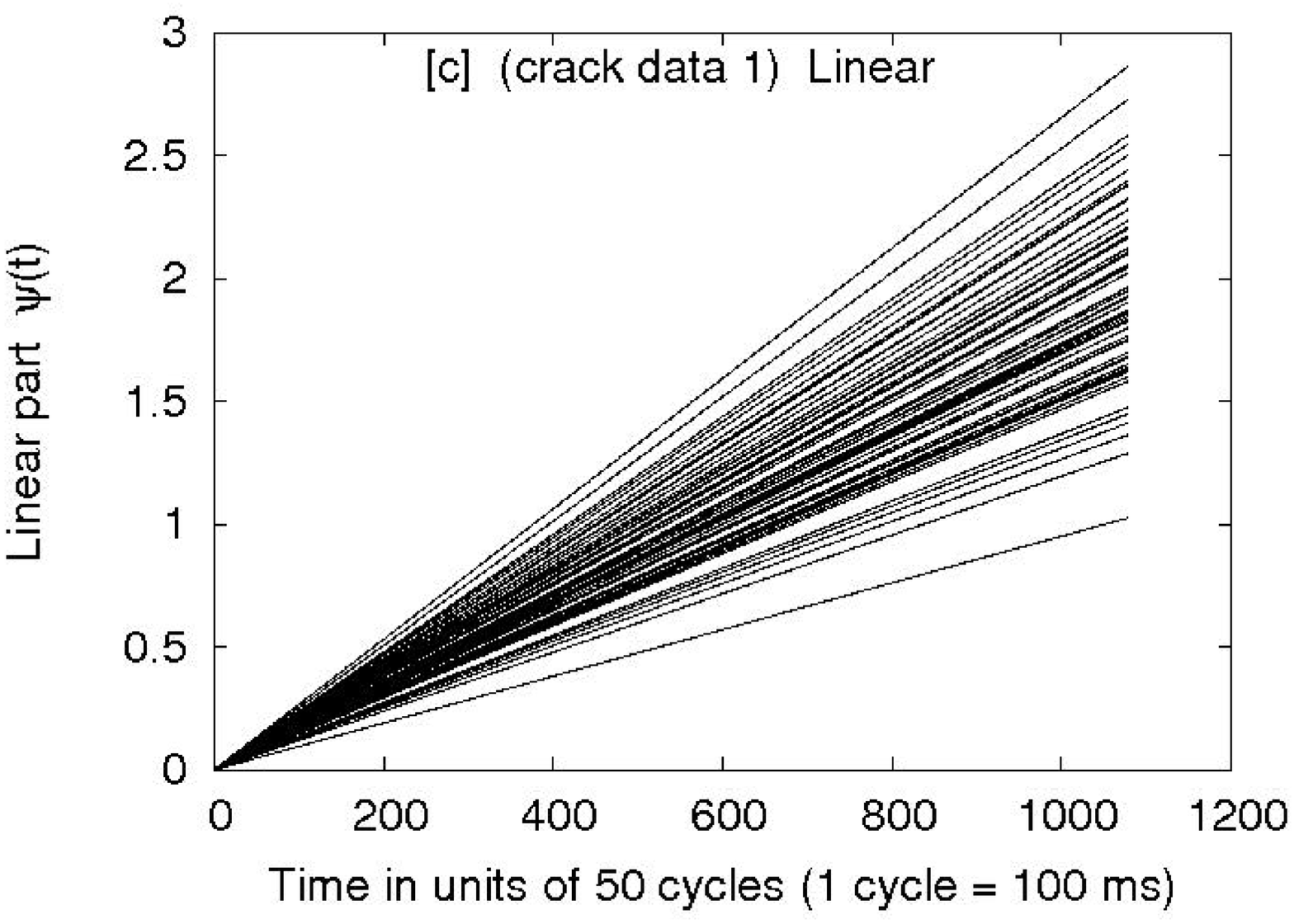,height=10cm,width=7.0cm,angle=-90}
\caption{ [a] Curves $\psi(t)$ obtained with Eq. (\ref{Omega5})
for the experimental data of 7075-T6 aluminum alloy for set \#1.
The value of $m$ used is $m=4.0$. [b] Principal part of the K-L
decomposition against [c] the linear approximation of the
continuous model made in Eqs. (\ref{eq1aa99}) plus (\ref{eq1aa9})
of the curves $\psi(t)$.  }
\end{figure}

\newpage
\begin{figure}[tbp]
\epsfig{file=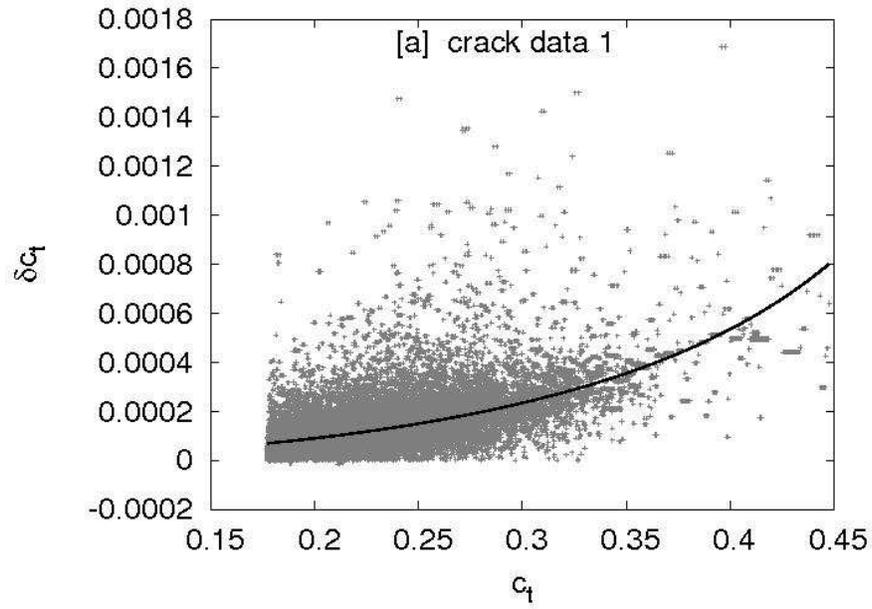,height=13cm,width=10.0cm,angle=-90} %
\caption{ Increments $\delta c_t$ against crack length $c_t$ fit
with Eq. (\ref {stochastic}) (solid curve) for the experimental
data of 7075-T6 aluminum alloy for the set \#1: $\Omega ~\Delta
S^{m}=0.0019\pm 0.0002$ and $m=4.0\pm0.2 $. }
\end{figure}

\newpage
\begin{figure}[tbp]
\epsfig{file=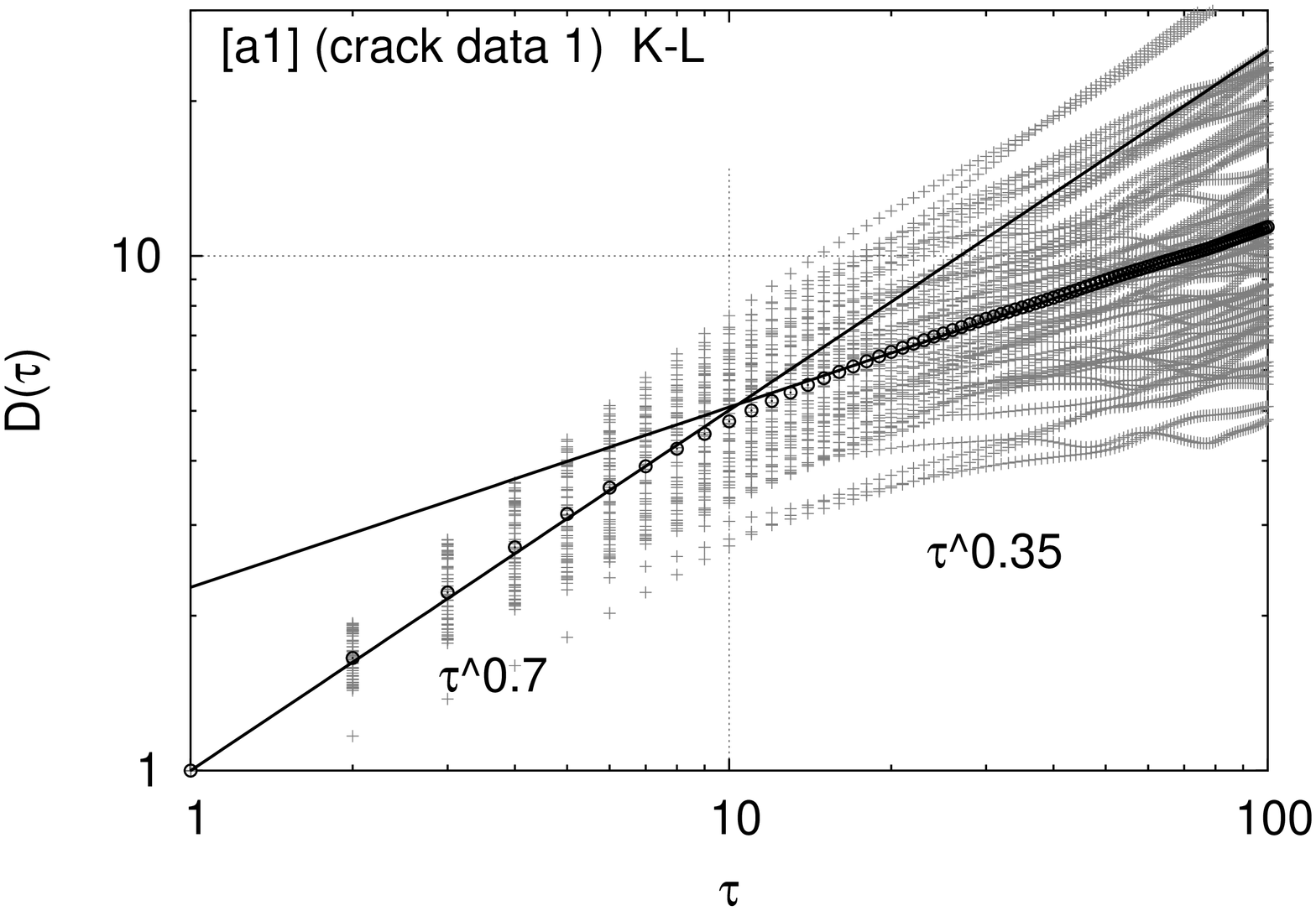,height=7cm,width=4.5cm,angle=-90} %
\epsfig{file=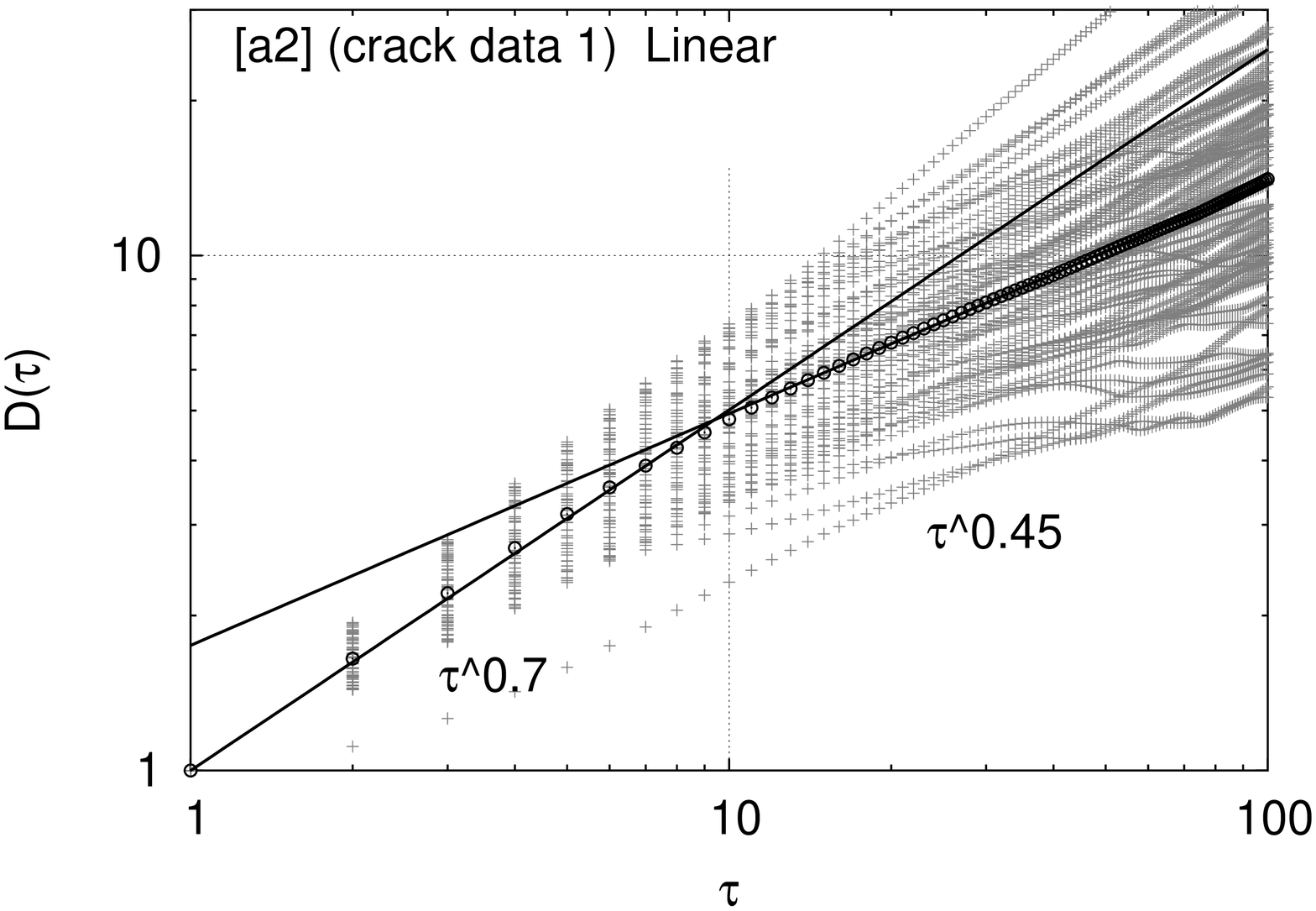,height=7cm,width=4.5cm,angle=-90}
\par
\epsfig{file=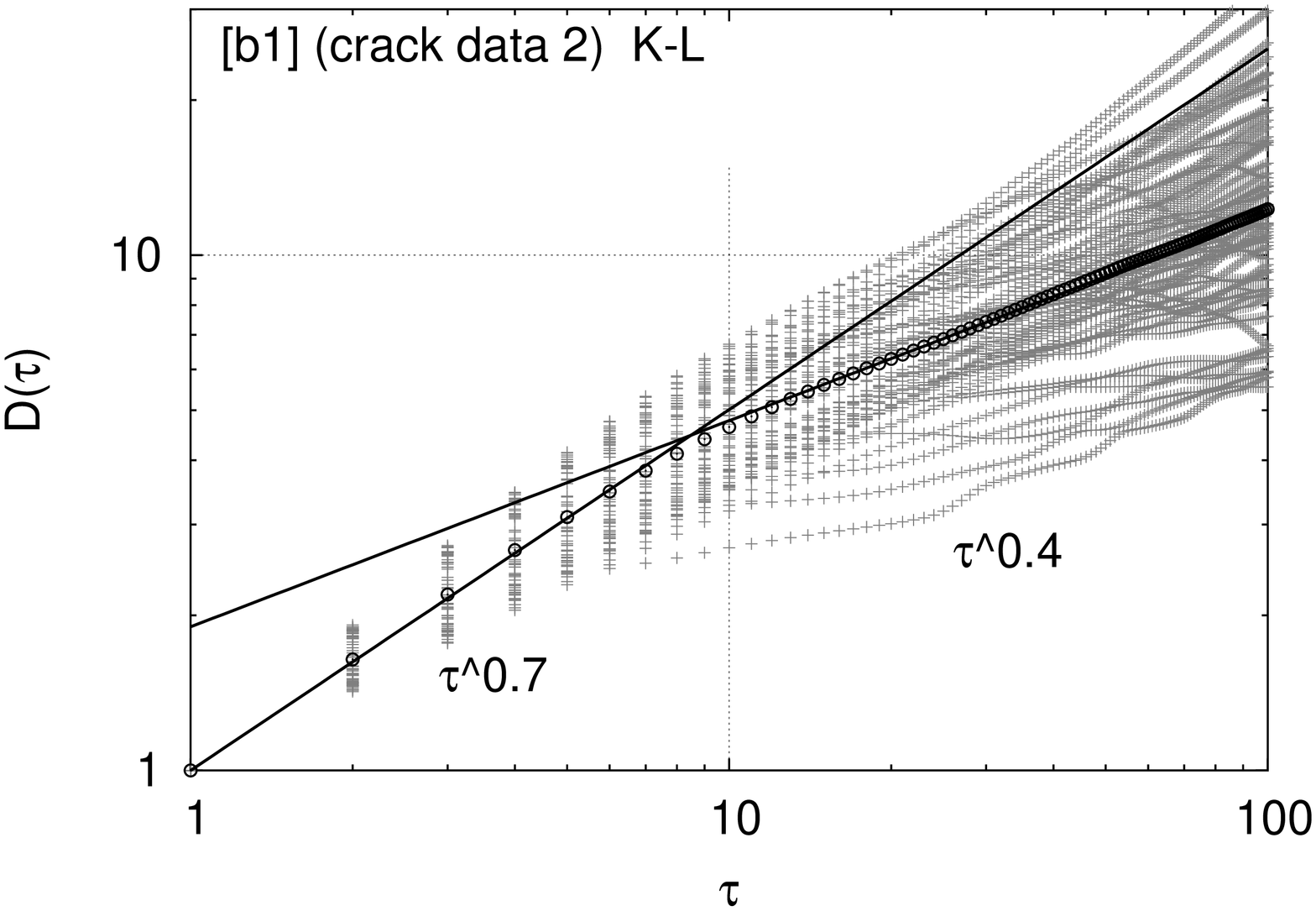,height=7cm,width=4.5cm,angle=-90} %
\epsfig{file=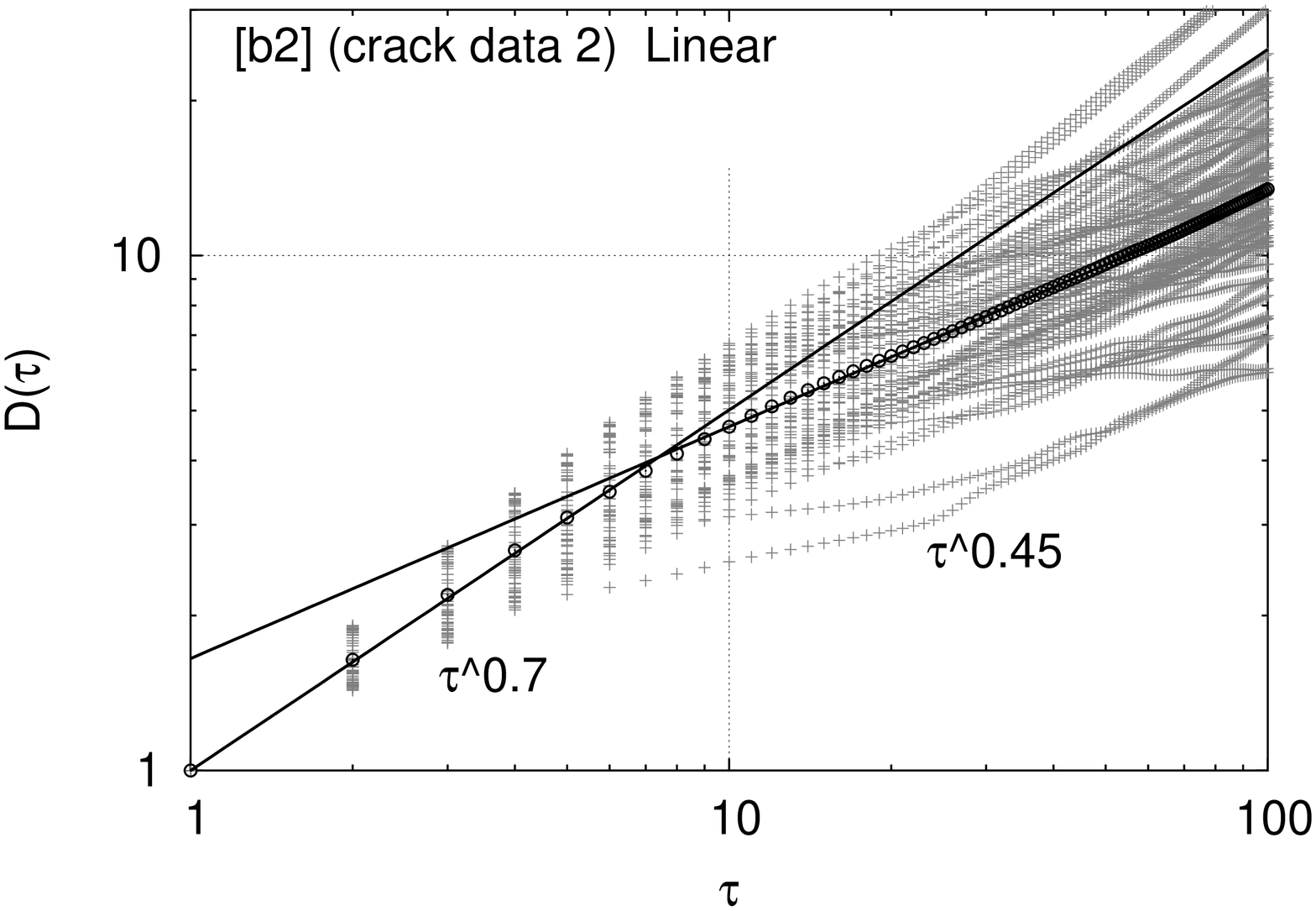,height=7cm,width=4.5cm,angle=-90}
\par
\epsfig{file=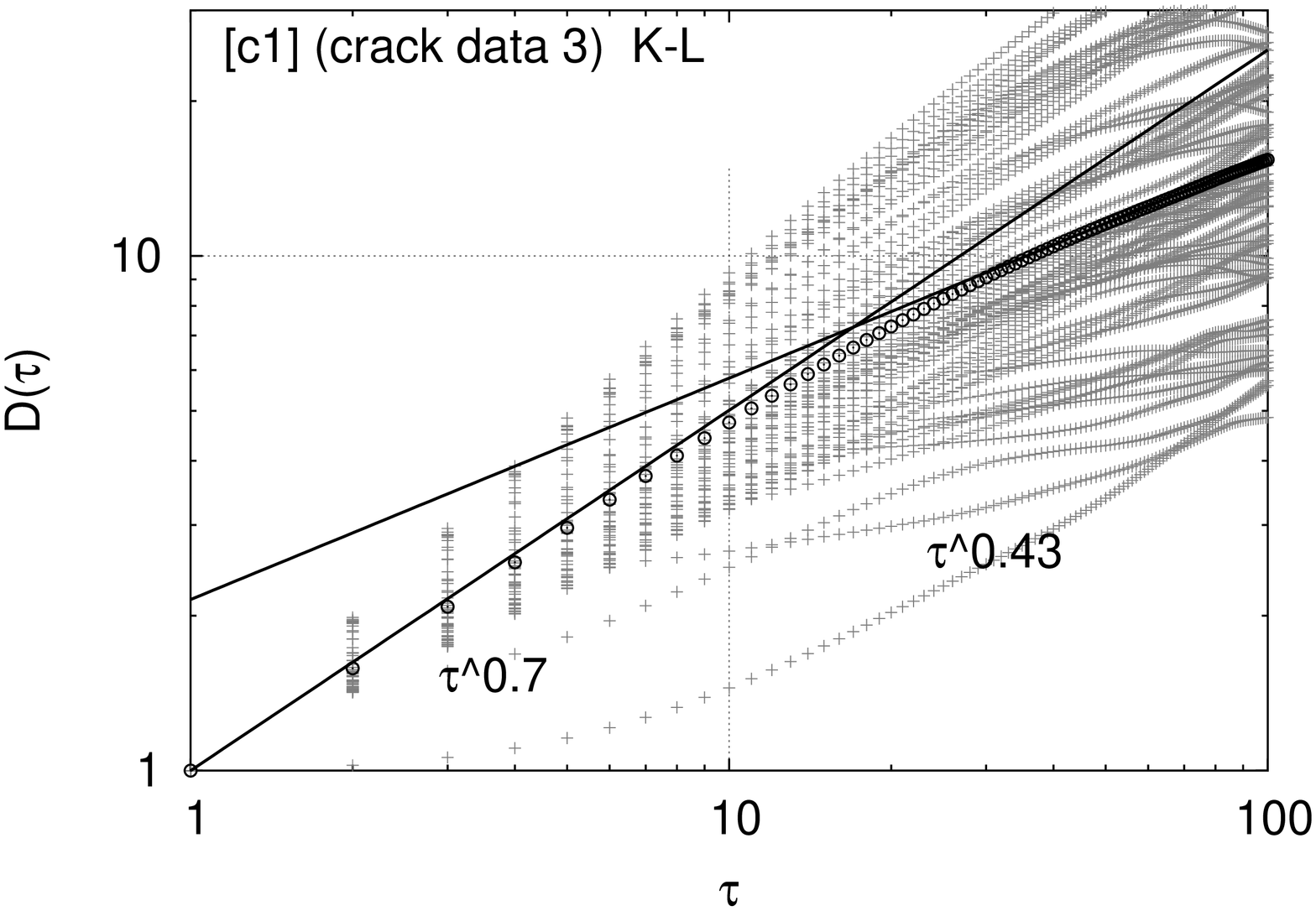,height=7cm,width=4.5cm,angle=-90} %
\epsfig{file=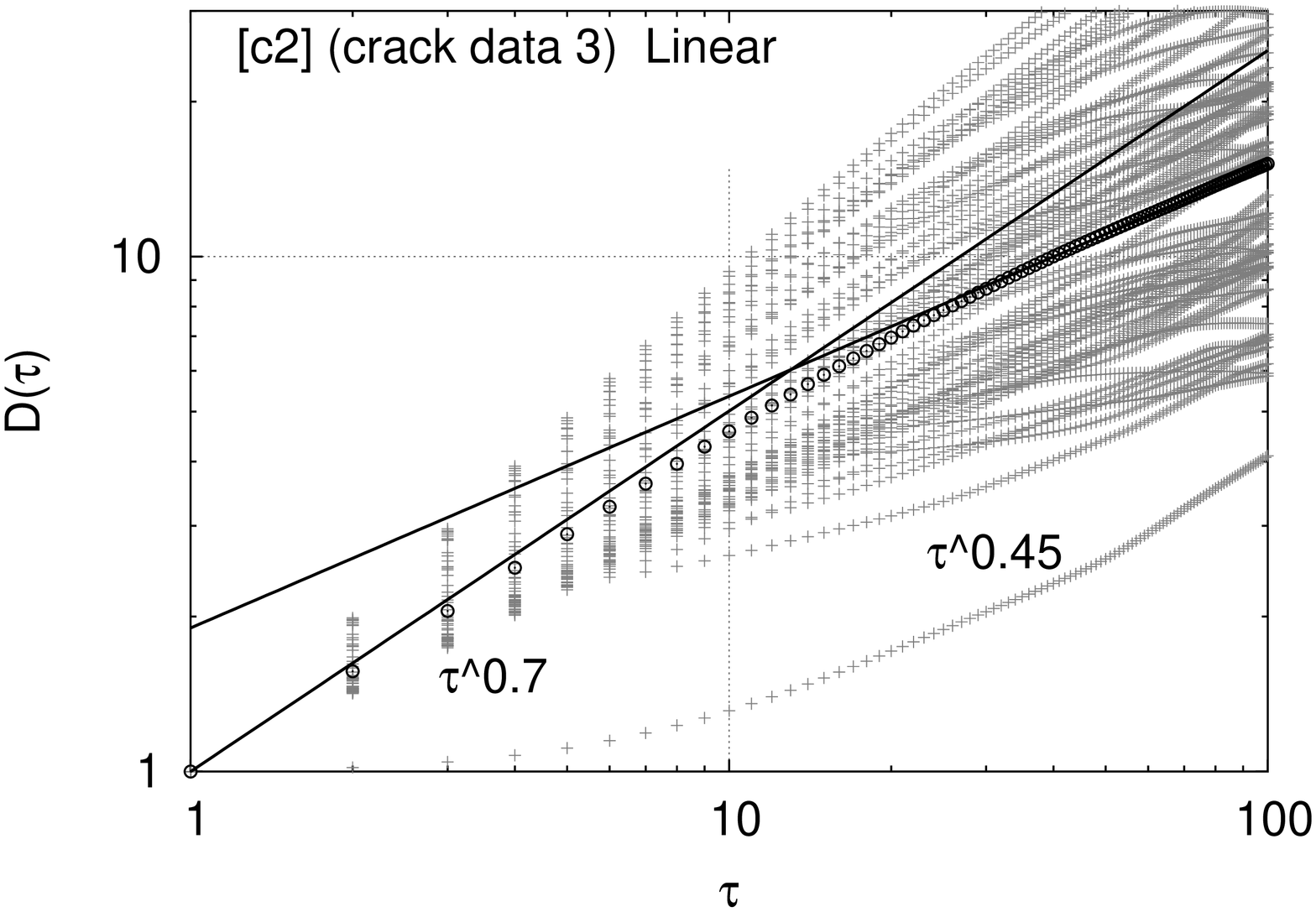,height=7cm,width=4.5cm,angle=-90} \caption{
 SDA for the residual part of the KL-decomposition (left figures) against SDA for
 the residual part of
linear approximation (right figures) of the continuous model. Note
the scaling
  transition at $\tau \approx 10$ from $H\approx 0.7$ to $H\approx 0.5$
   in both cases for all datasets.}
\end{figure}

\newpage
\begin{figure}[tbp]
\epsfig{file=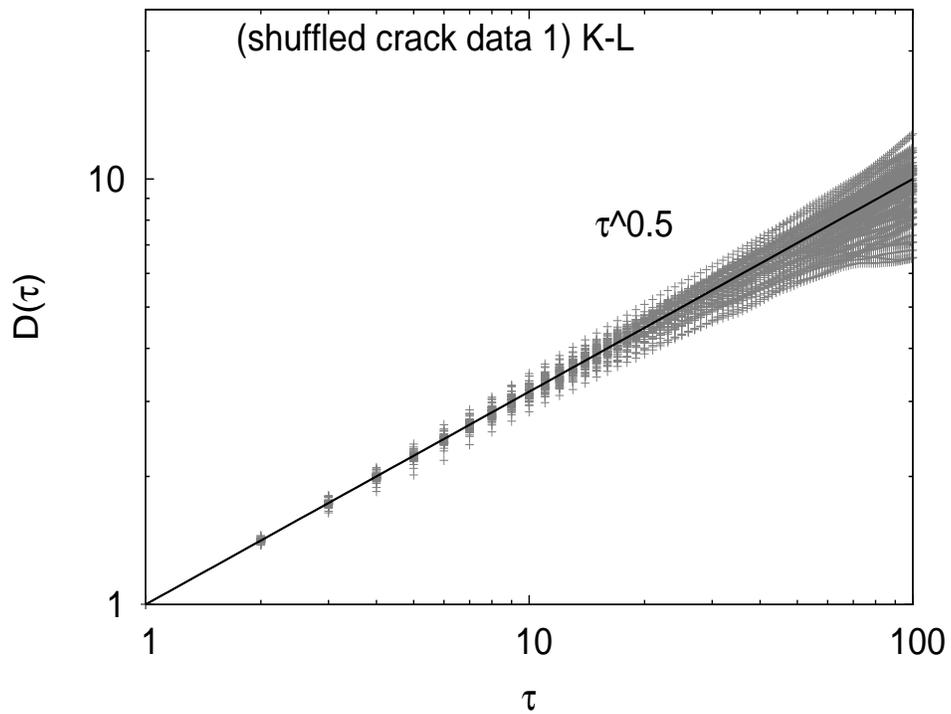,height=13cm,width=10.0cm,angle=-90} %
 \caption{
 SDA for the residual part of the KL-decomposition after shuffling of the increments $\{\theta_t\}$. Note
the random scaling
 of  $H\approx 0.5$.  The data refer to set \#1 and the
 comparison has to be made with Fig. 4 (crack data 1) KL.}
\end{figure}

\newpage
\begin{figure}[tbp]
\epsfig{file=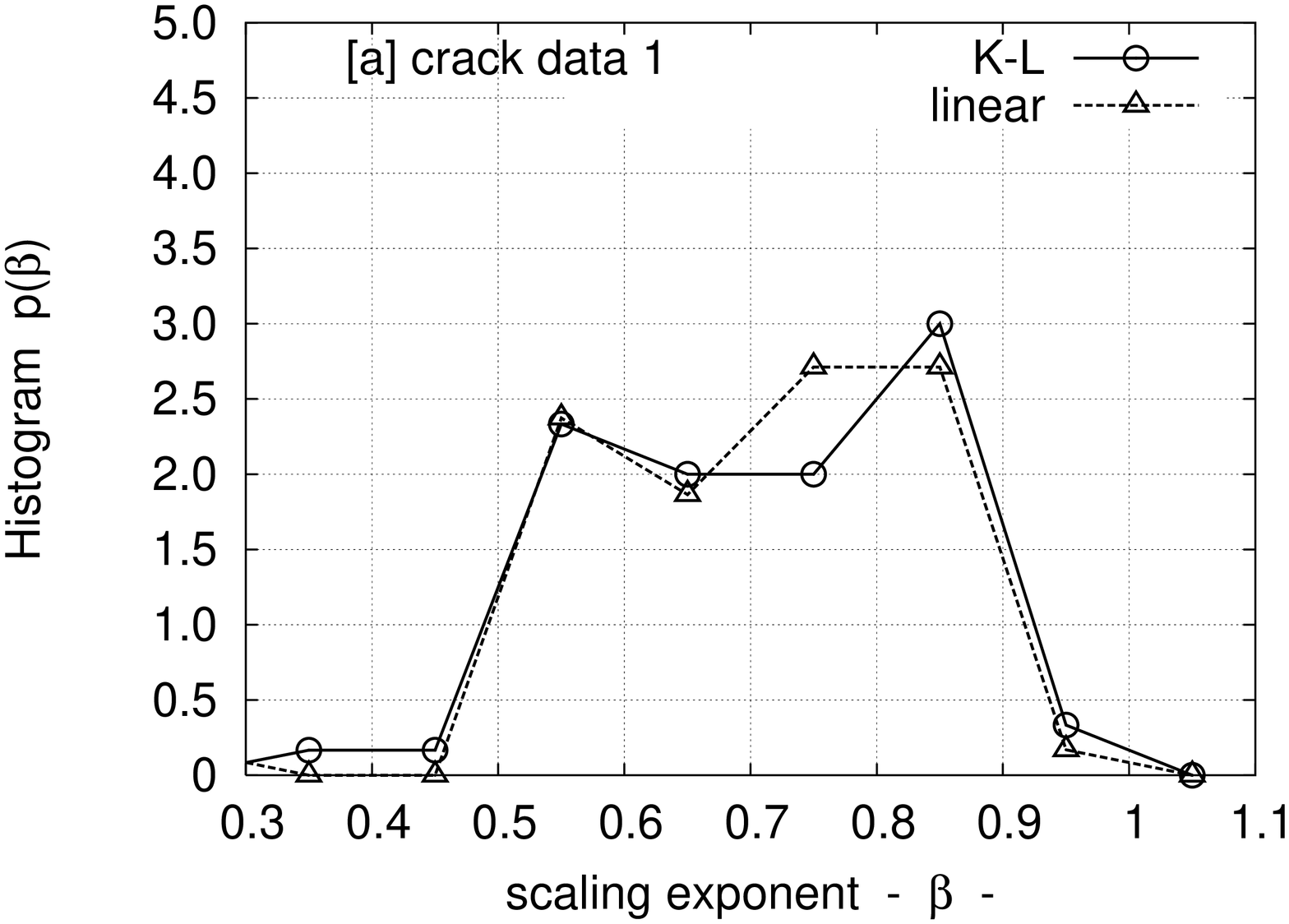,height=7cm,width=4.5cm,angle=-90} %
\epsfig{file=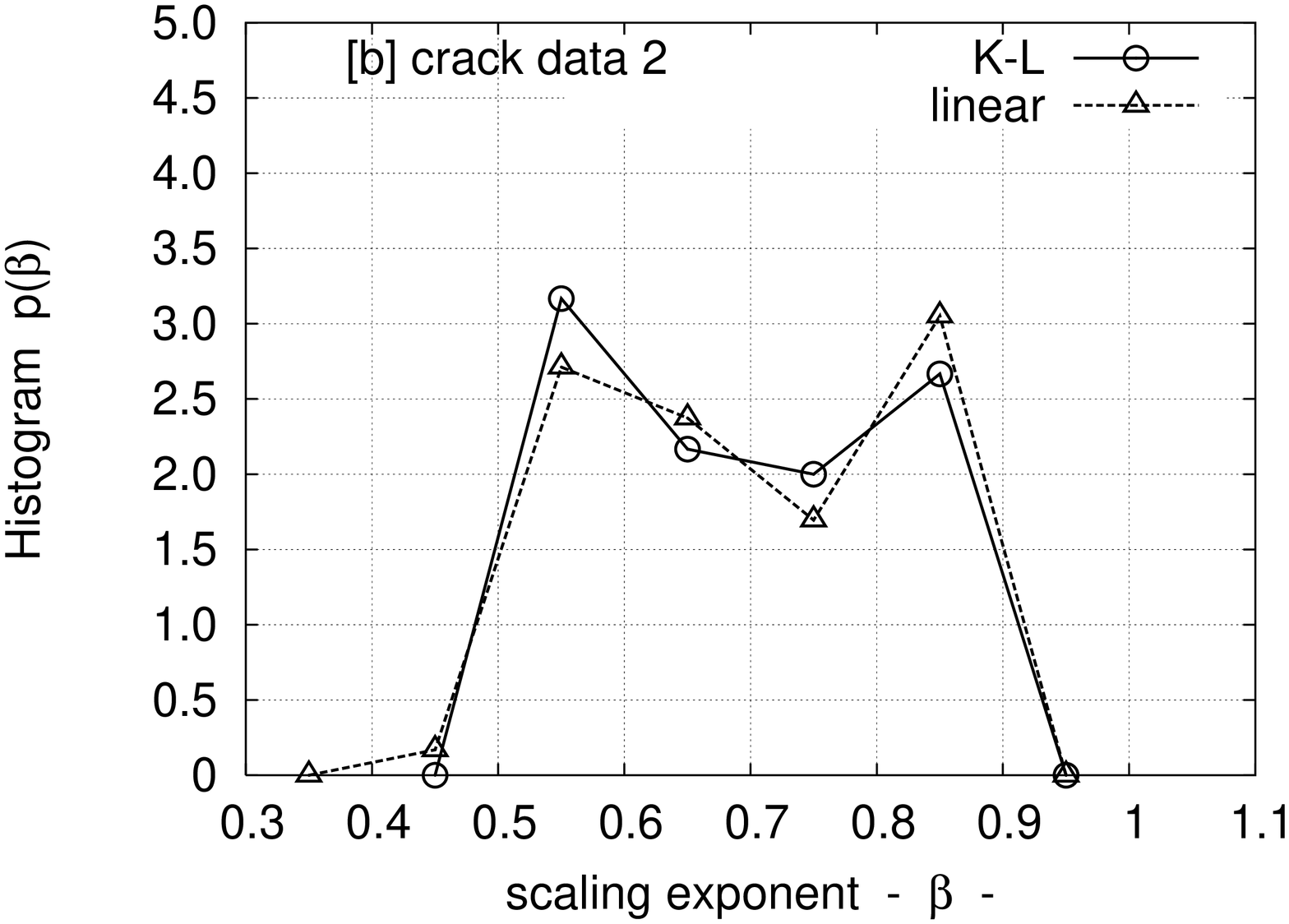,height=7cm,width=4.5cm,angle=-90}
\par
\epsfig{file=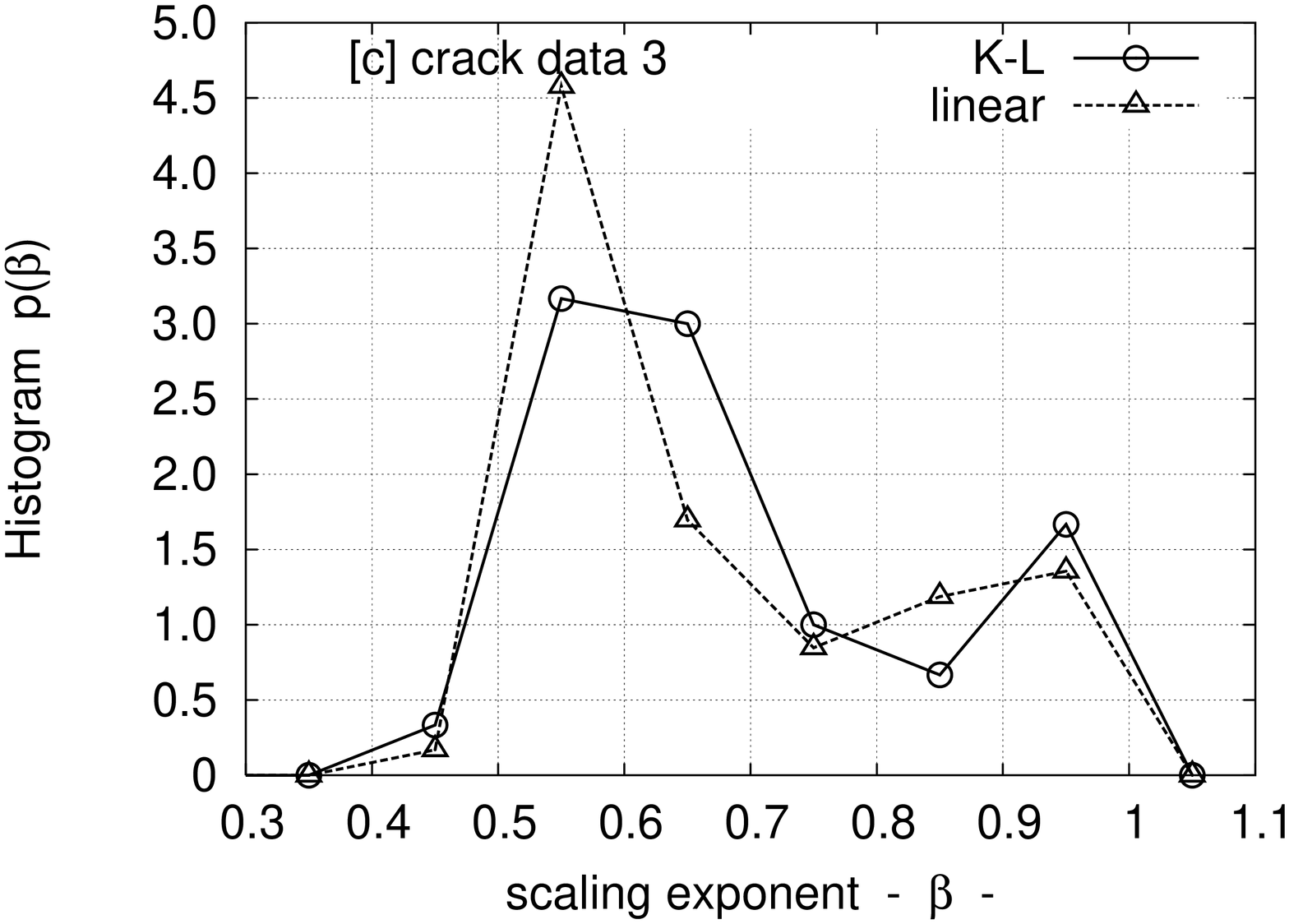,height=7cm,width=4.5cm,angle=-90} %
\epsfig{file=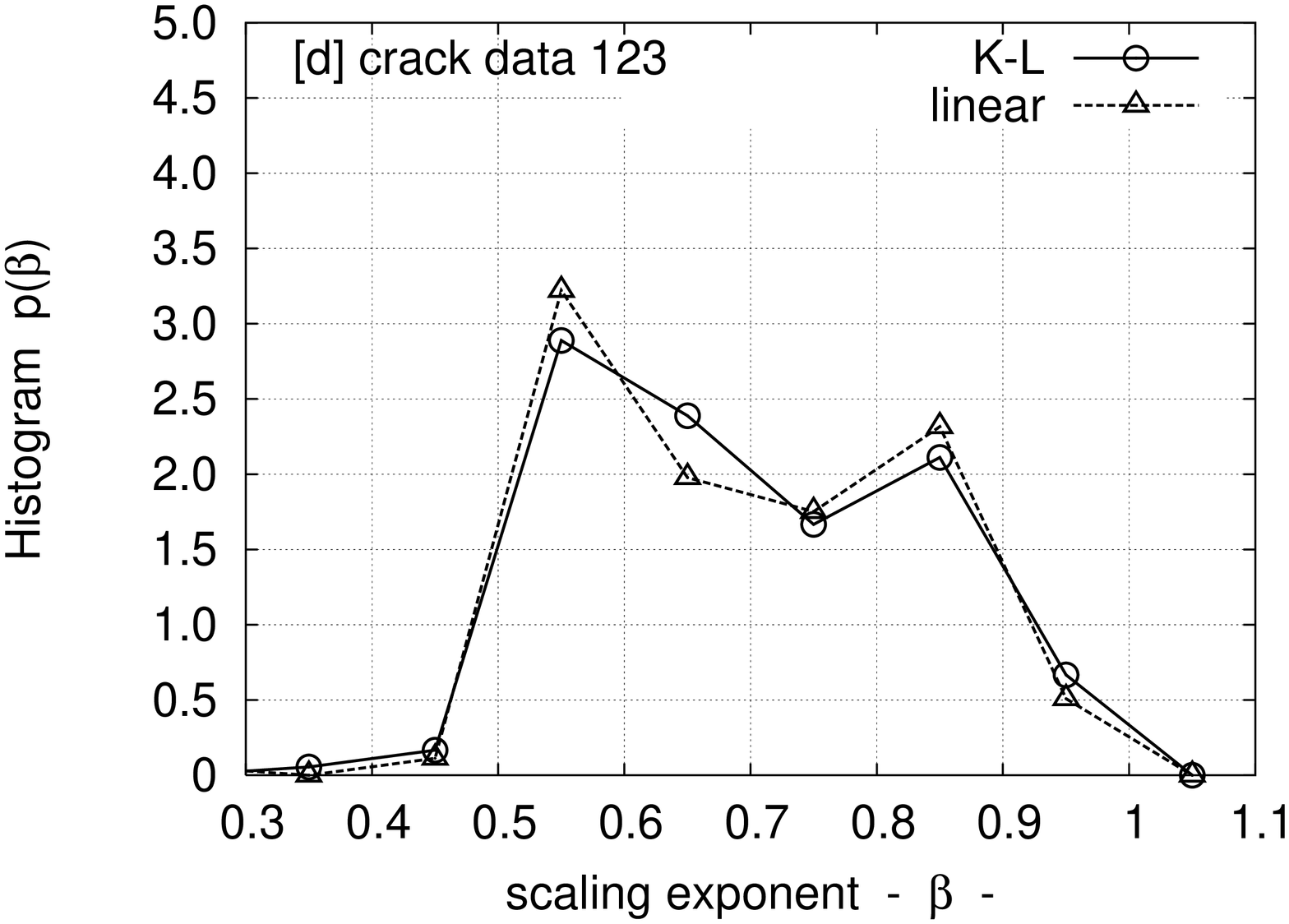,height=7cm,width=4.5cm,angle=-90} \caption{
Histogram and probability density function of the scaling exponent
$\beta$ estimated by fitting the the range $1<\tau<=10$ for each
curve shown in the plates of Figure 4. Each plate compares the
distributions obtained with the K-L decomposition and the continuous
linear model for each of the three crack data sets. Figure 7d shows
the histograms of all data. }
\end{figure}

\newpage
\begin{figure}[tbp]
\epsfig{file=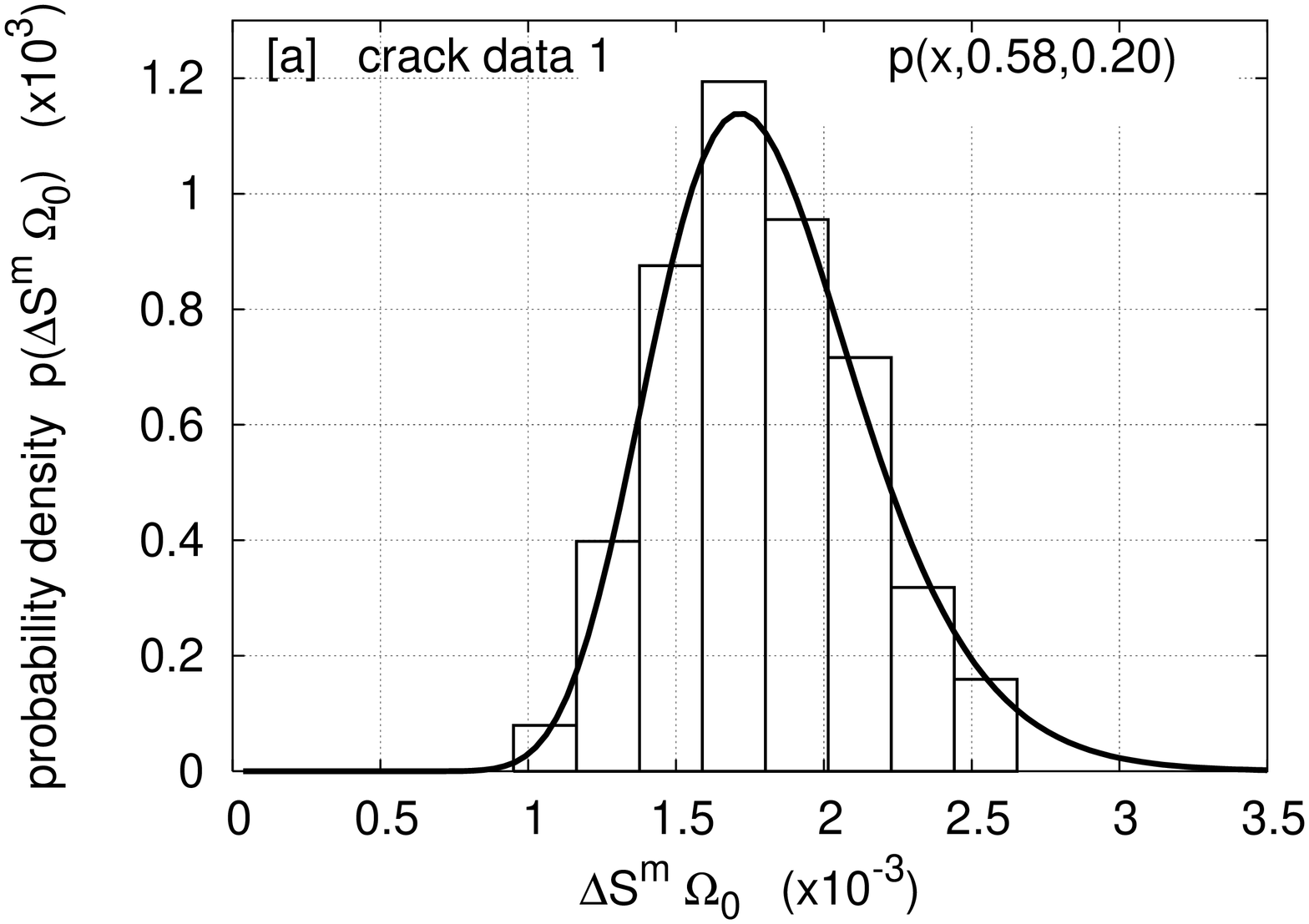,height=10cm,width=7.0cm,angle=-90} %
\epsfig{file=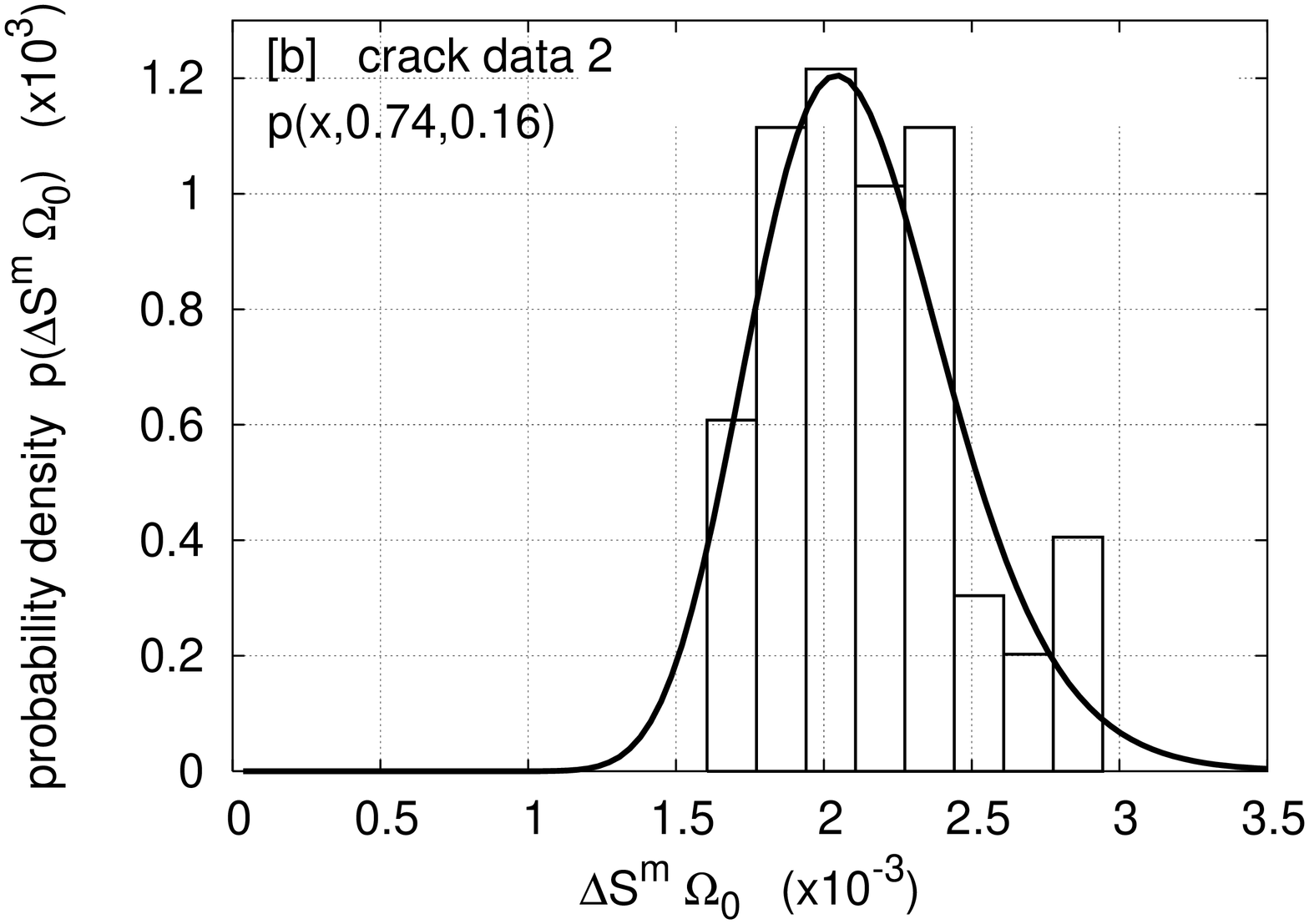,height=10cm,width=7.0cm,angle=-90} %
\epsfig{file=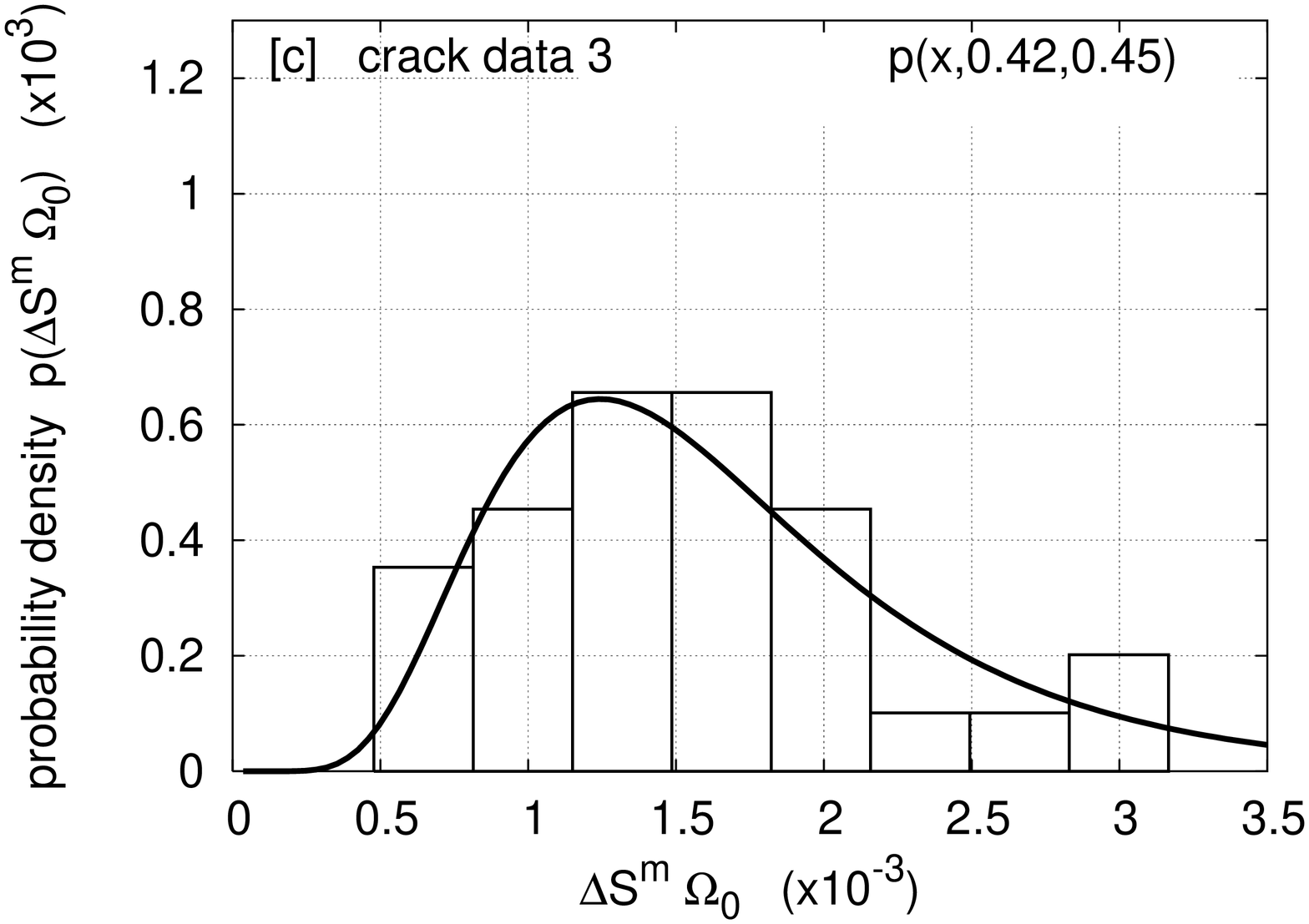,height=10cm,width=7.0cm,angle=-90}
\caption{ Histogram of the quantities $\Delta S^m~\Omega_0$ of the
continuous model of the experimental data presented by Eq.
(\ref{eq19}). The histograms are fitted with a lognormal
distribution $p(x;\mu,\sigma)$ shown in Eq. (\ref{lognor}). }
\end{figure}

\newpage
\begin{figure}[tbp]
\epsfig{file=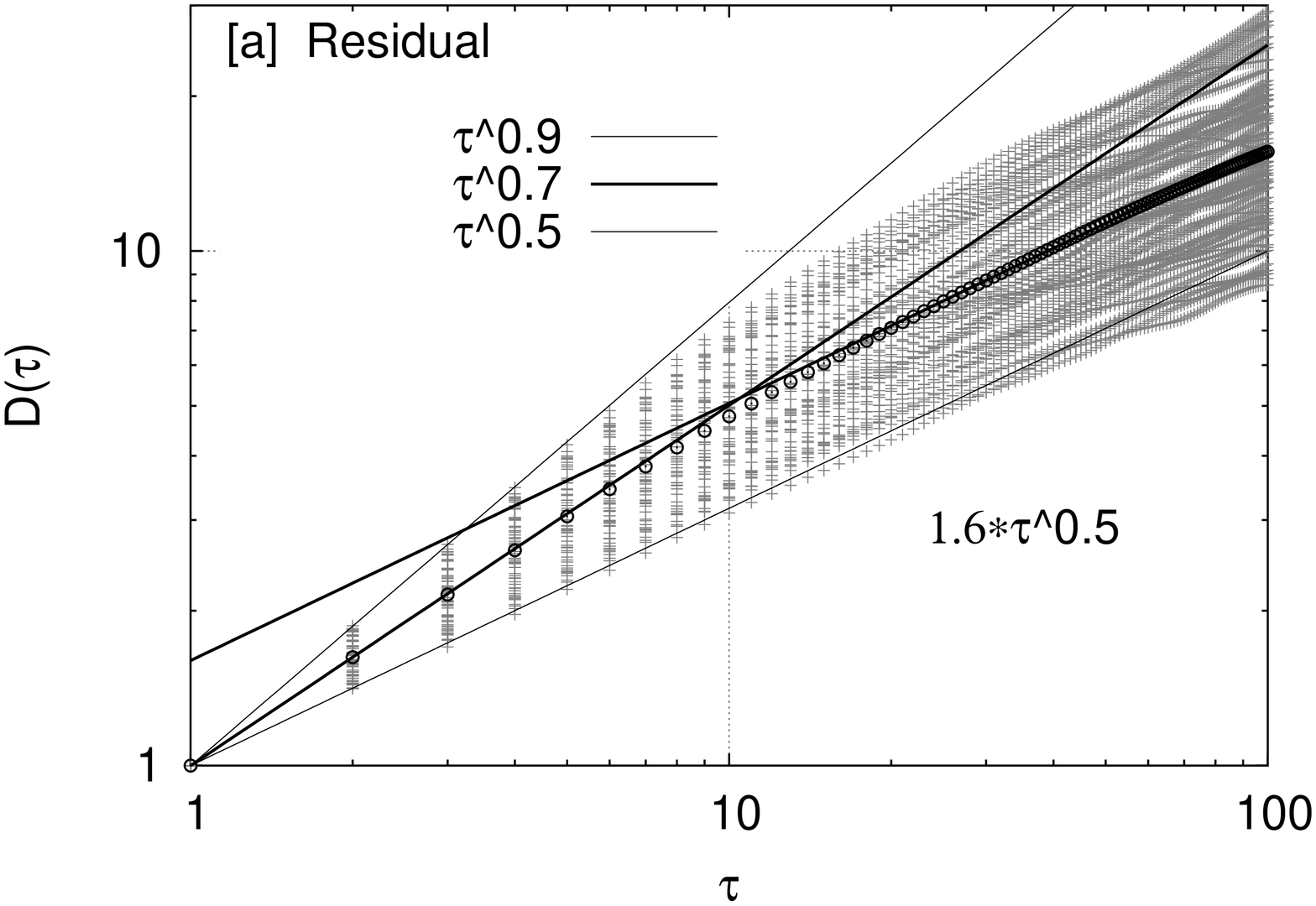,height=8cm,width=6.0cm,angle=-90} %
\epsfig{file=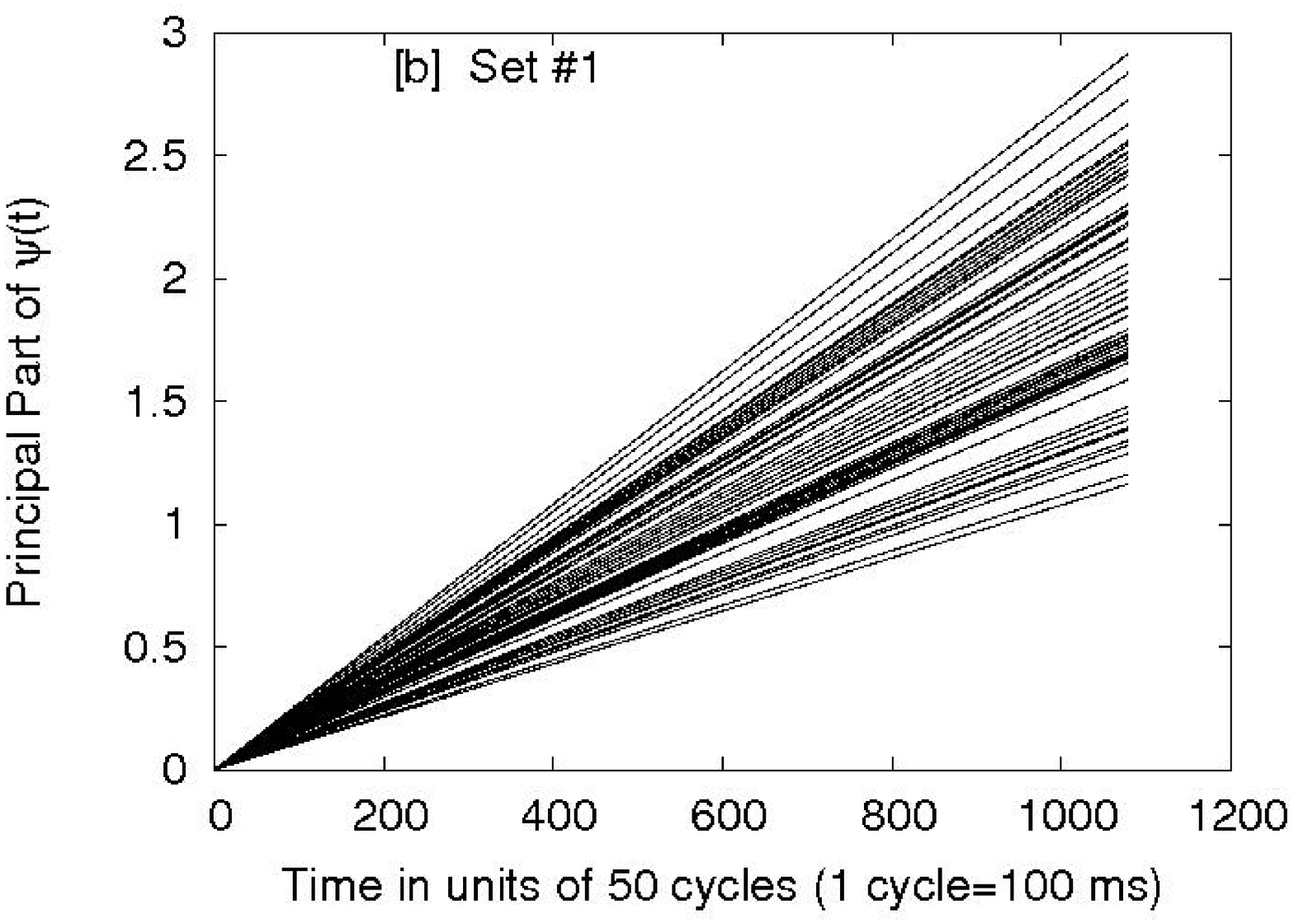,height=8cm,width=6.0cm,angle=-90}
\par
\epsfig{file=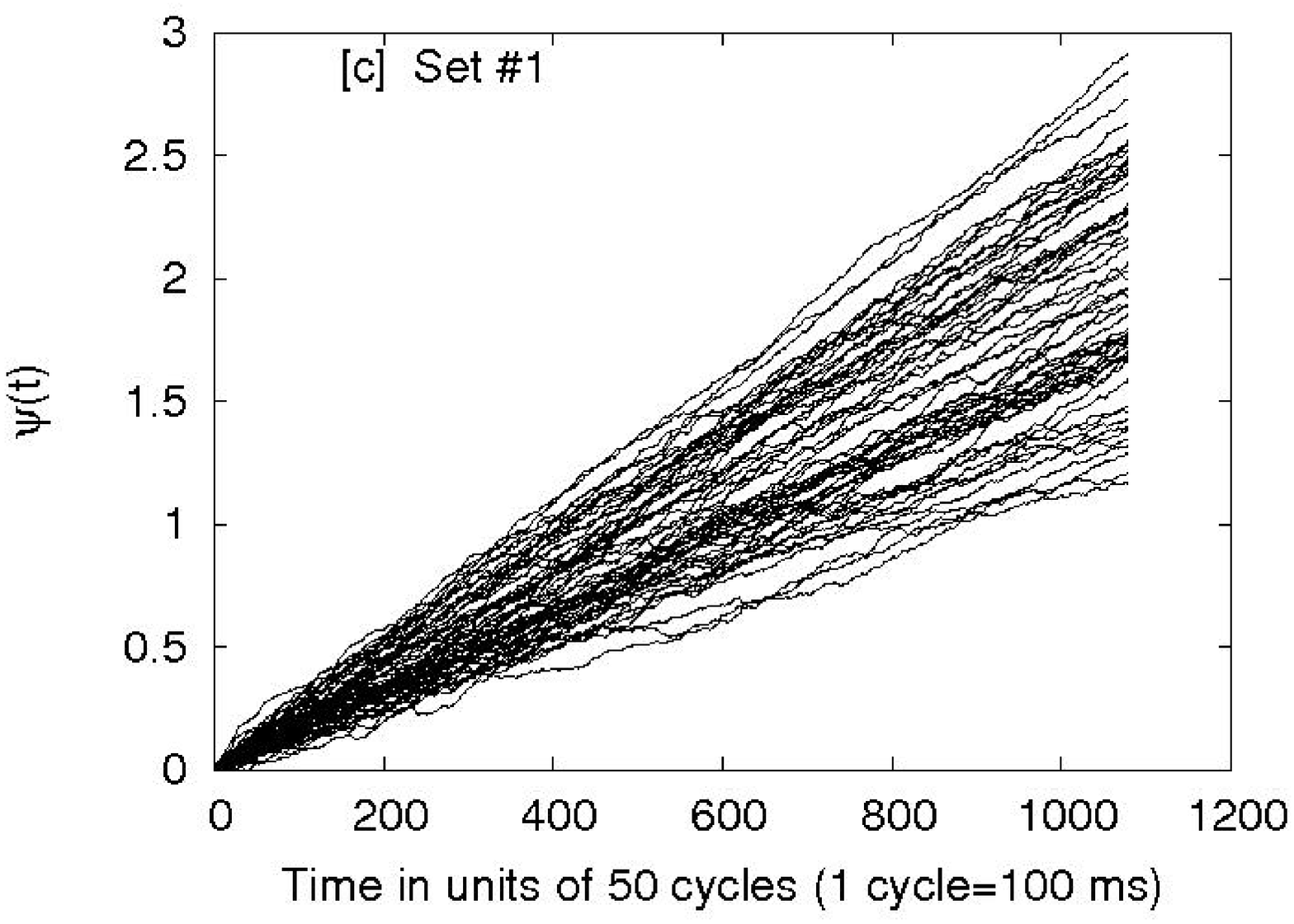,height=8cm,width=6.0cm,angle=-90} %
\epsfig{file=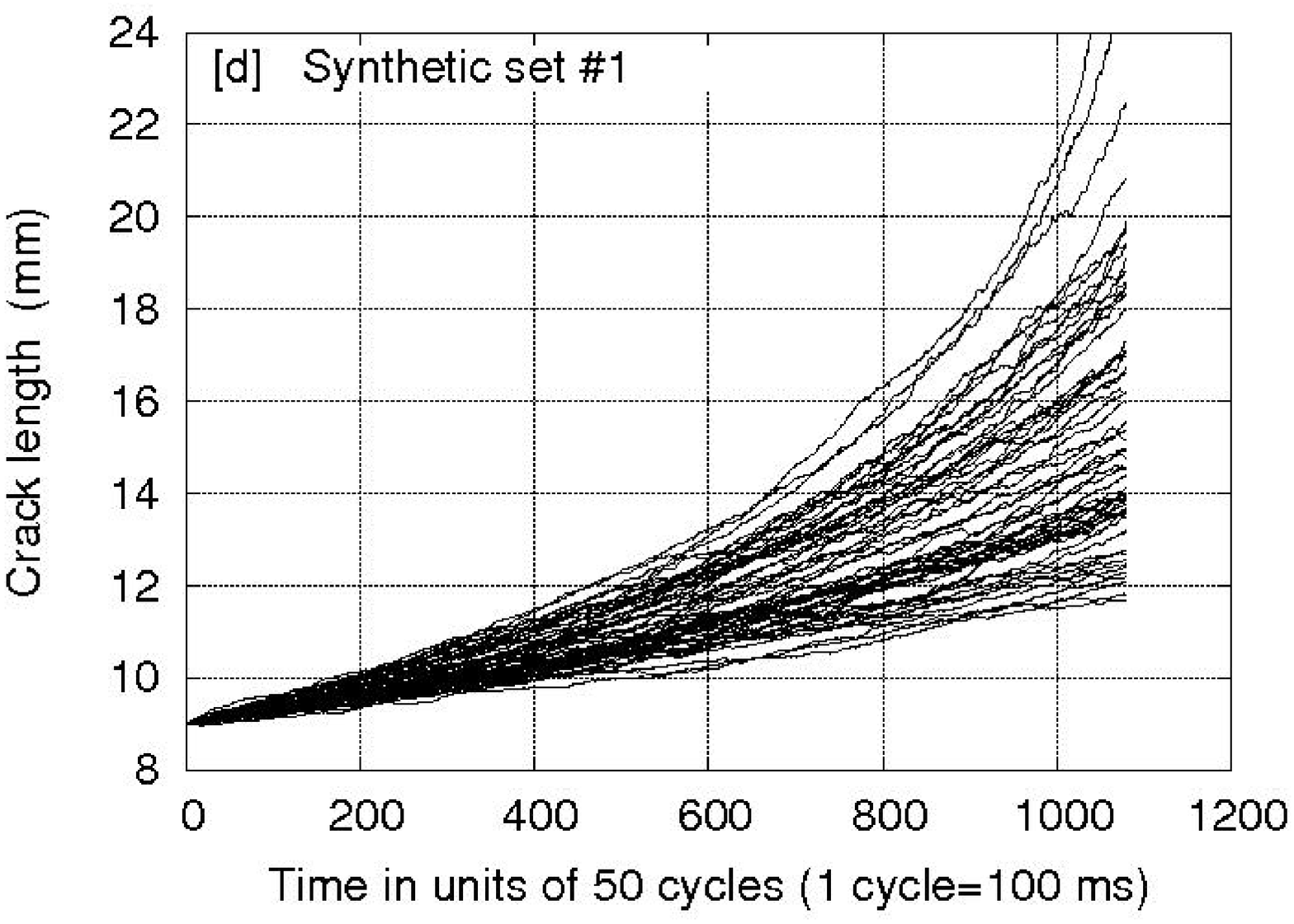,height=8cm,width=6.0cm,angle=-90} \caption{
Synthetic data of crack length for Set \#1. [a] SDA of the residual
component; [b] ballistic growth; [c] damage increment $\psi(t)$; [d]
crack length, compare with Figure 1a. }
\end{figure}

\end{document}